\documentclass{aa}  
\usepackage{graphicx}
\usepackage{txfonts}
\usepackage{aalongtable}
\begin{document}
\title{The Light Element Abundance Distribution in NGC~5128 from Planetary Nebulae\footnote{
Based on observations collected at the European Organisation for Astronomical 
Research in the Southern Hemisphere, Chile in observing proposals 64.N-0219, 66.B-0134, 67.B-0111 and 71.B-0134.}
}
\titlerunning{Abundance Distribution in NGC~5128 from PNe}

\author{
J. R. Walsh\inst{1}
\and
G. H. Jacoby\inst{2}
\and
R. F. Peletier\inst{3}
\and
N. A. Walton\inst{4}
}

\offprints{J. R. Walsh}

\institute{European Southern Observatory, Karl-Schwarzschild-Strasse 2, 
D-85748 Garching, Germany \\
\email{jwalsh@eso.org}
\and
Giant Magellan Telescope Organization / Carnegie Observatories,
Pasadena, USA \\
\email{gjacoby@gmto.org}
\and
Kapteyn Astronomical Institute, Postbus 800, 9700 AV Groningen, 
The Netherlands \\
\email{r.f.peletier@astro.rug.nl}
\and
Institute of Astronomy, University of Cambridge, Madingley Road, 
Cambridge CB3 0HA, UK
\email{naw@ast.cam.ac.uk}
}
\date{Received ; accepted}
\authorrunning{Walsh, J. R., et al.}
% \abstract{}{}{}{}{} 
% 5 {} token are mandatory
 
\abstract
{Planetary nebulae in the nearest large elliptical galaxy
provide light element abundances difficult or impossible to 
measure by other means in a stellar system very different 
from the galaxies in the Local Group.}
{The light element abundance pattern from many planetary nebulae (PNe) at a range 
of radial distances was measured from optical spectroscopy in the 
elliptical galaxy NGC~5128, which hosts the radio source Centaurus A. 
The PN abundances, in particular for oxygen, and the PN progenitor 
properties are related to the galaxy stellar properties.} 
{PNe in NGC~5128 covering the upper 4 mag. of the luminosity function 
were selected from a catalogue. VLT FORS1 multi-slit spectra in 
blue and red ranges were obtained over three fields at 3, 9 and 
15$'$ projected radii (4, 8 and 17 kpc, for an adopted distance of 
3.8 Mpc) and spectra were extracted for 51 PNe. 

Accurate electron temperature and density diagnostics are usually 
required for abundance determination, but were not available for most 
of the PNe. Cloudy photoionization models were run to match the spectra 
by a spherical, constant density nebula ionized by a black body central
star. He, N, O and Ne abundances with respect to H were determined and, for
brighter PN, S and Ar; central star luminosities and temperatures are
also derived.}
{
Emission line ratios for the 51 PNe are entirely typical of PN such 
as in the Milky Way. The temperature sensitive [O~III]4363\AA\ line was 
weakly detected in 10 PNe, both [O~II] and [O~III] lines were detected in 30
PNe, and only the bright [O~III]5007\AA\ line was detected in 7 PN. For 40 
PNe with Cloudy models, from the upper 2 mag. of the [O~III] luminosity 
function, the most reliably estimated element, oxygen, has a mean 
12+Log(O/H) of 8.52 with a narrow distribution. No obvious radial gradient 
is apparent in O/H over a range 2-20 kpc. Comparison of the PN
abundances with the stellar population, from the spectra of the  
integrated stellar light on the multi-slits and existing photometric 
studies, suggests an average metallicity of [Fe/H]=-0.4 and [O/Fe]=0.25.
}
{The masses of the PN central stars in NGC~5128 deduced from model tracks 
imply an epoch of formation even more recent than found for the minority young 
population from colour magnitude studies. The PN may belong to the young tail 
of a recent, minor, star formation episode or derive from other evolutionary 
channels, perhaps involving binary stars.}


\maketitle
%
%________________________________________________________________

\section{Introduction}
Planetary nebulae provide a multi-facetted probe of the galactic
environment. As a short-lived product of the evolution of low
mass stars, they sample the bulk, by both number and mass, of
the stellar population of galaxies of all Hubble types. Their
size makes them unresolved from the ground at distances greater
than $\sim$1 Mpc but their strong emission lines makes them 
easy targets for observational work, even when observed 
against the high stellar background of a galaxy bulge. These
advantages have been exploited in a number of distinct areas.
Statistics of PN in galaxies of various types
show a late time indicator of the star formation 
rate through  the luminosity specific PN frequency $\phi$ (
Ciardullo et al. \cite{Ciardullo89}). The PN luminosity function (PNLF), 
despite theoretical difficulties, still proves to be a powerful 
distance method (Jacoby et al. \cite{Jacoby92}; Ciardullo et al
\cite{Ciardullo02}). The radial velocity, simply measured from one
emission line, provides a kinematic test particle for study of
the galaxy potential and in intra-cluster gas of the cluster
potential and turbulence (Arnaboldi et al. \cite{Arnaboldi}). 
The abundances of a number of light elements can be measured 
from emission line ratios in PN spectra, such as He and N
and, in particular, the $\alpha$-elements 
O, Ne, S and Ar, which are difficult to study in 
all but high resolution and high signal-to-noise spectra of 
individual stars.

As the closest example of a large early-type galaxy, 
NGC~5128 (Hubble type S0p) occupies a central place in 
studies of resolved stellar populations in a galaxy 
different to the Milky Way and M31. This large 
elliptical galaxy in the Centaurus group shows 
signs of major activity with an active nucleus, 
Faranoff-Riley Class I radio lobes, presence of dust and 
young stars in its inner region (Graham \cite{Graham}), a young
($\sim$0.3 Gyr) tidal stream (Peng et al. \cite{Peng02}) and
stellar shells in its outer regions (Malin et al. \cite{Malin});
see Israel (\cite{Israel}) for an earlier review. This 
activity can be attributed to a minor merger which
has had little influence on the bulk of the older passively
evolving population (Woodley \cite{Woodley06}), making NGC~5128 a
nearby exemplar of more distant massive early-type galaxies;
Harris (\cite{Harris10}) provides a recent review of the underlying 
galaxy properties. The spectroscopy of globular clusters by Beasley et
al. (\cite{Beasley}) indicates typical ages of 7-8 Gyr 
from stellar population fitting, reinforcing the presence of
a large-scale intermediate age star formation episode. Fits
to deep colour-magnitude diagrams also provide evidence for
a minority, much younger, population of age 2-4 Gyr (Rejkuba et
al. \cite{Rejkuba11}). 

As representative of the low mass stars, planetary 
nebulae in NGC~5128 have been catalogued over a period
of two decades, beginning with the catalogue of Hui et al. 
(\cite{Hui93a}). 785 PN were discovered by [O III]5007\AA\ 
emission line and off-band imaging; from the PN [O~III] 
magnitudes Hui et al. (\cite{Hui93b}) 
determined a PNLF distance of 3.5 Mpc. Independent 
measurements from the Mira period-luminosity relation 
and the luminosity of the tip of the red giant branch
(Rejkuba et al. \cite{Rejkuba05}), surface brightness
fluctuations (Tonry et al. \cite{Tonry}), globular cluster 
luminosity function (Harris et al. \cite{Harris88}) and 42 
classical Cepheid variables (Ferrarese et al. \cite{Ferrarese}) 
result in distance estimate in the range 3.4 to 4.1 Mpc. 
Harris et al. (\cite{HRH10})  present a comprehensive review 
of distance estimates to NGC~5128, and recommend a best-estimate
of 3.8 Mpc, which is adopted here (distance modulus 
27.90mag.). Peng et al. (\cite{Peng04}) extended the original catalogue
of PN and found a further 356 PN by filter imaging; 780 
out of a total of 1141 were spectroscopically confirmed.
Further emission line mapping and follow-up intermediate
dispersion spectroscopy has extended this list to over
1200 confirmed PN (Rejkuba \& Walsh \cite{RejWal}). This number
of PN makes NGC~5128 a rich source for statistical
extra-galactic PN studies, rivalled only by the Milky Way and M~31
(Merrett et al. \cite{Merrett06}).

Since there are both a large number of PN and they have 
been catalogued to large galactocentric distances
(to 80 kpc along the major axis by Peng et al. (\cite{Peng04})), 
the radial velocities allow the dynamics of the halo mass 
distribution to be studied. Hui et al. (\cite{Hui95}) measured
[O~III] radial velocities for 431 out of their 785 
catalogued PN. The offset of the rotation axis from the
minor axis was attributed to evidence of triaxiality; fitting the
rotation curve and velocity dispersion revealed a
radially increasing mass-to-light ratio and hence
the presence of a dark matter halo. Peng et al. (\cite{Peng04}) 
extended the PN kinematic work to 780 PN and showed 
the large rotation along the major axis but with a 
pronounced zero-velocity twist produced by the 
triaxial-prolate mass distribution. The kinematics of
the large population of globular clusters (with 563 available 
radial velocities -- Woodley et al. \cite{Woodley10}, Woodley 
et al. \cite{Woodley07}, Beasley et al. 
\cite{Beasley}) show similar kinematics to the PN but with 
small differences, such as lower rotation amplitude (Woodley
et al. \cite{Woodley07}). 

The NGC~5128 globular clusters (GCs) also provide 
fundamental evidence of the star formation history.
Of the 605 confirmed GCs in NGC~5128 
(Woodley et al. \cite{Woodley07}, Woodley et al. \cite{Woodley10}),
more than half have metallicity measurements and they divide 
roughly half and half into metal rich and metal
poor above and below [Fe/H]$=-1.0$. The metal poor
GCs ([Z/H]$\sim-1.3$) are older with ages similar to 
Milky Way GCs and the metal rich GCs (Peng et al. \cite{Peng04};
Beasley et al. \cite{Beasley}) have intermediate ages (4-8 Gyr)
and metallicity ([Z/H]$\sim-0.5$). There are many
more GC candidates (e.g. Harris et al. \cite{Harris04}) and the
total number of GCs has been estimated at around 1300 
(Harris \cite{Harris10}). 

NGC~5128 is close enough that individual stars can
be resolved with HST imaging (and from the ground with
adaptive optics in the near-IR) and several studies
have been obtained outside of the bright
bulge where crowding is lower. Soria et al. (\cite{Soria})
detected the Red Giant Branch (RGB) and the Asymptotic
Giant Branch (AGB) from intermediate age stars. The same field
was followed up with NICMOS photometry and the IR colour
magnitude diagram shows stars well above the tip of the
red giant branch for old stars, confirming the presence
of intermediate age stars (Marleau et al. \cite{Marleau}). Colour magnitude
diagrams from HST WFPC2 were constructed for two halo fields at 
projected radii of 19 and 29 kpc (Harris et al. \cite{Harris99}; 
Harris et al. \cite{Harris00}). The colour magnitude 
diagrams are dominated by old stars. Under the assumption that the 
ages of the stars in NGC~5128 are similar to those of old globular 
clusters in our Galaxy, the mean metallicity of the two fields
was found to be similar at [Fe/H]$\sim-0.4$ 
with about one third of the stars in a metal poor 
component and the rest metal rich. The metallicities 
in the halo fields were compared by Harris et al. 
(\cite{Harris02}) to the metallity distribution function 
(MDF) in a field at 7.4kpc, containing a mixture of outer 
bulge and inner halo. A broad MDF was found as in the halo 
fields, peaking at [Fe/H]$\sim-0.4$; 
but subtracting the MDF of the halo fields
reveals a peak at slightly higher metallicity ($\sim-0.2$).
Much deeper ACS imaging by Rejkuba et al. (\cite{Rejkuba05}) in a
halo field at 36 kpc reached to
the horizontal branch (core helium burning population)
as well as showing the red giant branch, red clump and 
AGB bump. Again a broad MDF was found
with an average metallicity of $-0.64$, but with a broad tail to
higher metallicity ($>0$). The age sensitive indicators imply
an average age for the halo of 8$^{+3}_{-4}$ Gyr. Comparison 
of the MDF in the four fields shows the peak shifting to 
lower metallicity with increasing projected radius but the 
distribution is similarly wide at all radii.

In order to determine light element abundances of PN in
NGC~5128 fairly high signal-to-noise spectroscopy is 
required. Close to the inner bulge the stellar continuum 
surface brightness is too high and only the strongest
few lines can be measured. However in the outer regions by 
selecting bright PN, diagnostic
lines of He, Ne, Ar and S as well as the brighter lines of 
O and N can be measured and abundances derived. Walsh
et al. (\cite{Walsh99}) performed deep ESO 3.6m long slit spectra of 
a few selected PN with a long slit spectrograph and 
measured lines in addition to the brightest line
([O III]5007\AA) in five PN. O/H could be determined
in two PN and values of 12 + Log(O/H) $\sim$8.5 
(i.e. [O/H] - 0.2, adopting the Solar oxygen abundance of 8.69 
from Scott et al. (\cite{Scott})) were derived. This initial
work has been expanded to deeper spectra of many more PNe with 
the ESO VLT and FORS1 instrument in multi-slit mode. In section 2 
the observations are described and the reduction of the spectra and 
derivation of the abundances are presented in section 3. The results 
for the PN line fluxes and derived abundances both for
the individual PN and for summations of many spectra 
are presented in section 4. Since the diagnostic lines 
for electron temperature and density, important for
abundance determination, are weak or undetected, 
photoionization modelling of all well-detected lines was
undertaken using Cloudy and is described in section 5.
Consideration of the lack of a gradient in the PN O
abundances, comparison of the results with the 
stellar, photometrically-derived metallicity, and 
relation of the PN to the stellar populations 
in NGC~5128 are discussed in section 6. Conclusions 
are collected in section 7.

\section{Observations}

\subsection{FORS imaging and spectroscopy}

%ADD
The European Southern Observatory Very Large Telescope (VLT) 
FORS1 instrument mounted on VLT Unit Telescope 1 was used for
the observations described. FORS1 has a number of modes for imaging; 
for spectroscopy, long slit, multi-slit, or multi-slit mask are 
available (see Appenzeller et al. \cite{Appenzeller} for details).
The multi-object mode (MOS) allows 19 slitlets of height 
varying from 20 to 22 $\arcsec$ to be placed over a 
6.8$\times$6.8$\arcmin$ field of view. In the regions of 
NGC~5128 where PNe were catalogued by Hui et al (\cite{Hui93a}) and
(\cite{Hui93b}), the surface
density is such that there are usually enough PNe to be able to fill
the slitlets, thus ensuring an optimal match between 
target density and instrument multiplex. This match was
realised in practice for regions over the bright disk of NGC~5128,
whilst in the outer regions, around 50\% of the slitlets
could be utilised. Three regions were selected for MOS spectroscopy
to fulfill the following criteria: sample the inner and outer
regions at a range of effective radii to detect any abundance
gradient from the PNe; sample the major and minor axes; ensure
that some of the brightest PNe were observed to maximize the
probability of detecting the faintest line species; collect as
many PN spectra as possible. Not all of these criteria are
mutually consistent, but the three regions selected at
radial offset distances of 3.4, 8.7 and 15.4$\arcmin$ provided
spectra of 51 PNe at a range of m$_{5007A}$ (
m$_{5007A}$ = -2.5 log F$_\lambda$ (erg cm$^{-2}$ s$^{-1}$) -13.74;
Jacoby \cite{Jacoby89}) from 23.5 to $>$27.0 mag.

\subsubsection{Pre-imaging}
Direct images with FORS1 and a narrow band [O~III] filter (called
OIII+50, centred at 5005\AA\ and FWHM 57\AA) were obtained in December 
1999-January 2000\footnote{ESO programme 64.N-0219} in service mode.
Tab. \ref{PreImaObs} lists details of the imaging observations whose primary
purpose was to provide images for the MOS slit assignment using
the FIMS tool. The standard resolution 
collimator was used giving
a field of 6.8$\times$6.8$\arcmin$ with a pixel size of 0.2$\arcsec$ 
per pixel. The field names are taken from those of Hui
et al. (\cite{Hui93a}). A companion
filter (actually a redshifted [O~III] filter
OIII/6000, centred at 5109~\AA with FWHM 61\AA) was also used to obtain 
an [O~III] off-band subtraction to confirm the reality of the PNe; 
all the previously catalogued PNe in the fields were confirmed as 
[O~III] emission line objects. Seeing at the time of observations was 
between 0.5 and 1.2$\arcsec$.

\begin{table*}
\caption[]{FORS1 pre-imaging observations}
\label{PreImaObs}
\centering 
\begin{tabular}{l l l l c} 
\hline\hline  
Position & ~~~~~~~$\alpha$ ~~~~~~~~~~~~ $\delta$ & [O~III] Exp & Cont. & Date \\
(radius, kpc)  &  ~h ~~m ~~s ~~~~~~$^\circ$ ~~~$\arcmin$ ~~~$\arcsec$    &   (sec)     &  (sec) &     \\
\hline  \\
F56 (9.6) & 13 25 50.6 -43 06 09.2 & 2$\times$480 & 2$\times$220 & 1999 12 27 \\
F42 (3.8) & 13 25 09.0 -43 02 29.3 & 2$\times$480 & 2$\times$220 & 2000 01 13 \\
F34 (17.0) & 13 24 47.2 -43 13 32.2 & 2$\times$480 & 2$\times$220 & 2000 01 13 \\
\hline
\end{tabular}
\end{table*}

\subsubsection{MOS Spectroscopy}
FORS1 MOS observations in service mode were conducted in three
sessions\footnote{ESO programmes 64.B-0219, 66.B-0134 and 
71.B-0134}. Tab. \ref{PNSpeObs} lists the salient details in chronological order. 
To cover the wavelength range with useful diagnostic emission 
lines (essentially from [O~II]3726,3729\AA\ to beyond [O~II]7320,7330\AA),
two grisms were employed: grism 600B (called GRIS\_600B+12) has a dispersion 
of 1.2\AA/pixel and, for a centred MOS slitlet, a wavelength coverage of 
3450 to 5900\AA; grism 300V (named GRIS\_300V+10 and used with a GG435 
blocking filter) has a lower dispersion of 2.7\AA/pix and a coverage of 
4450 to 8650\AA. The overlap region between both spectra is 
4500-5900\AA\, thus allowing at least the
strong [O~III]4959,5007\AA\ lines to be used to tie the two spectra
to a common flux scale. The slit width for the campaigns in 2000
and 2001 was 0.8$\arcsec$ whilst the
observations in 2003 had a slit width of 1.0$\arcsec$. The
resulting spectral resolutions are 4.8 and 6.0\AA\ for
600B and 10.7 and 13.4\AA\ for 300V observations respectively.
The supporting observations of the spectrophotometric standard stars 
for flux calibration, which were taken with a broad slit of 5$''$ width,
are listed in Tab. \ref{SpecSpeObs}.
 
\begin{table*}
\caption[]{FORS1 PN spectroscopic observations}
\label{PNSpeObs}
\centering 
\begin{tabular}{l l l l l l} 
\hline\hline  
Position & ~~~~~~~~$\alpha$ ~~~~~ $\delta$   & Grism &  ~~Exp.  & ~~Date & Seeing \\
         &  ~h ~~m ~~s ~~~~~~$^\circ$ ~~~$\arcmin$ ~~~$\arcsec$ &       & ~~(sec)  &      & ($\arcsec$) \\
\hline  \\
F56 & 13 25 50.6 -43 06 09.2 & 600B & 4$\times$2400 & 2000 05 02 & 0.8-1.2 \\
F56 & 13 25 50.6 -43 06 09.2 & 300V & 2$\times$2400 & 2000 05 02 & 0.8-1.2 \\
    &     &     &     &     & \\
F42 & 13 25 09.0 -43 02 29.3 & 600B & 2$\times$1500 & 2001 03 22 & 0.8 \\
F34 & 13 24 47.2 -43 13 32.2 & 600B & 2$\times$1500 & 2001 03 24 & 0.9 \\
F34 & 13 24 47.2 -43 13 32.2 & 600B & 2$\times$1500 & 2001 03 27 & 0.8 \\
F42 & 13 25 09.0 -43 02 29.3 & 600B & 2$\times$1500 & 2001 03 27 & 0.7 \\
     &     &     &     &     & \\
F34 & 13 24 47.2 -43 13 32.2 & 300V & 2$\times$1320 & 2003 03 25 & 0.7 \\
F34 & 13 24 47.2 -43 13 32.2 & 600B & 8$\times$1320 & 2003 04 08 & 0.5 \\
F42 & 13 25 09.0 -43 02 29.3 & 300V & 2$\times$1320 & 2003 04 24 & 1.0 \\
F42 & 13 25 09.0 -43 02 29.3 & 600B & 3$\times$1320 & 2003 04 24 & 1.0 \\
F42 & 13 25 09.0 -43 02 29.3 & 600B & 3$\times$1320 & 2003 04 30 & 0.6  \\
F42 & 13 25 09.0 -43 02 29.3 & 300V & 2$\times$1320 & 2003 04 30 & 0.5 \\
F56 & 13 25 50.6 -43 06 09.2 & 300V & 2$\times$1320 & 2003 05 05 & 0.5 \\
F56 & 13 25 50.6 -43 06 09.2 & 600B & 4$\times$1320 & 2003 05 05 & 0.6 \\
F34 & 13 24 47.2 -43 13 32.2 & 300V & 2$\times$1320 & 2003 06 02 & 1.1 \\
F56 & 13 25 50.6 -43 06 09.2 & 600B & 2$\times$1320 & 2003 07 21 & 1.0 \\
F56 & 13 25 50.6 -43 06 09.2 & 300V & 2$\times$1340 & 2003 07 22 & 0.4 \\
\hline
\end{tabular}
\end{table*}

\begin{table}
\caption[]{FORS1 spectrophotometric standard star observations}
\label{SpecSpeObs}
\centering 
\begin{tabular}{l l r l} 
\hline\hline  
Star & Grism &  Exp.  & Date \\
     &       & (sec)  &      \\
\hline  \\
LTT 7379 & 600B & 25 & 2000 05 02 \\
LTT 7379 & 300V & 7 & 2000 05 02 \\
         &      &     &            \\
LTT 7379 & 600B & 25 & 2001 03 24 \\
LTT 7379 & 600B & 25 & 2001 03 27 \\
         &      &     &            \\
EG 274   & 600B & 30 & 2003 04 08 \\
LTT 7379 & 600B & 25 & 2003 04 24 \\
LTT 7379 & 300V & 7 & 2003 04 24 \\
LTT 7379  & 600B & 25 & 2003 04 30 \\
LTT 7987 & 600B & 100 & 2003 05 05 \\
LTT 7987 & 300V & 30 & 2003 05 05 \\
LTT 7987 & 300V & 30 & 2003 06 02 \\
LTT 7379  & 300V & 7 & 2003 07 22 \\
\hline
\end{tabular}
\end{table}

\section{Reduction and analysis}

\begin{figure}
\centering
\resizebox{\hsize}{!}{\includegraphics{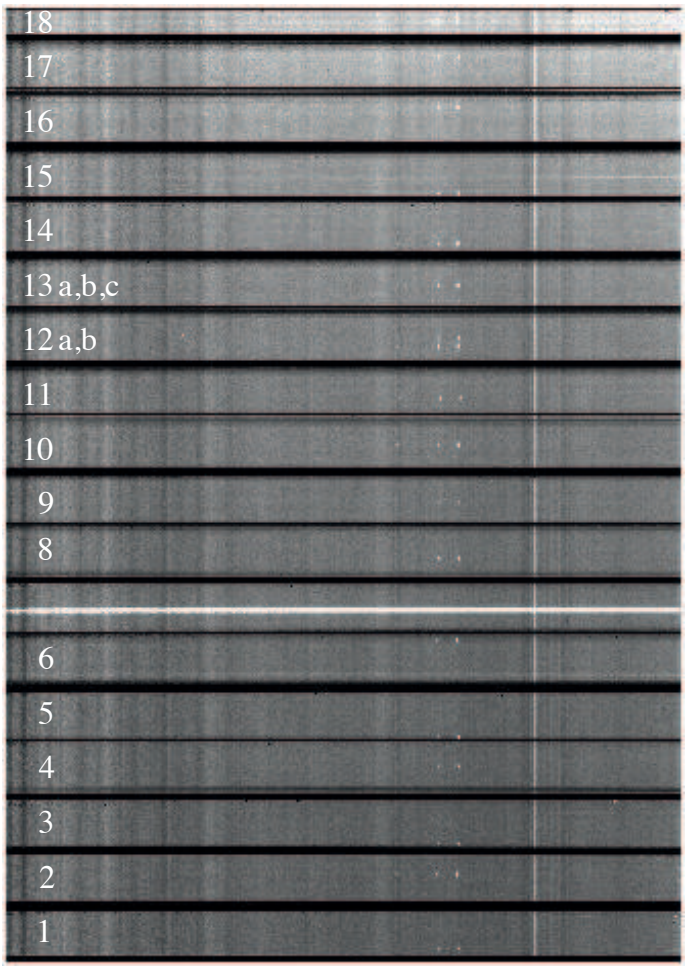}}
\caption{The reduced MOS spectra for field F56 taken with grism 
600B (data from May 2000). Eighteen of the 19 slitlets are shown for a wavelength 
range from about 3970 to 5520\AA\ and the numbering of the extracted 
PN spectra are indicated. Two slitlets show multiple PN spectra. 
The higher background stellar continuum towards the top of the image 
corresponds to lower galactocentric distance.}
\label{MOSspec}
\end{figure}

All the data frames (flats, science frames on PN and standard stars) 
were bias subtracted using master bias frames provided by the ESO
reduction pipeline contemporaneous with the observing data. Standard 
IRAF\footnote{IRAF is distributed by the National Optical Astronomy 
Observatories, which are operated by the Association of Universities 
for Research in Astronomy, Inc., under cooperative agreement with 
the National Science Foundation.} routines were used for the reduction. 
Pixel-to-pixel flat fields were constructed from dome flats by 
extracting the 2D area of each slitlet, collapsing in the 
cross-dispersion direction, box car smoothing in the dispersion 
direction and dividing the dome flat by the smoothed version. 
Identical reductions were performed for the bluer (600B grism) 
and redder (300V grism) spectra. Wavelength calibration was 
achieved by fitting 4th order Chebyshev polynomials to the 
dependence of pixel position on wavelength for the arc lamp 
lines on each slitlet separately. The separate slitlet
images, rebinned to an identical wavelength scale, were then 
recombined into a single image which was corrected for 
atmospheric extinction. Individual exposures were
combined with clipping to remove discrepant pixels caused by
detector bad pixels and cosmic ray events. Flux calibration 
was achieved by observations of one or more spectrophotometric 
standard stars, the ones applied being listed in Tab. 
\ref{SpecSpeObs}. The standard star spectra were analysed 
in an identical way to the PN, except that the one slitlet 
pertaining to the standard star was reduced. In the few 
cases where a standard star was not available in the same 
configuration as the PN on the night of 
observation, a standard star from another night had to be used. 
In general conditions were good or photometric, and no large flux 
calibration discrepancies were found between spectra of the same 
field observed in different runs (see Tab. \ref{PNSpeObs}). Narrow band 
magnitudes for the spectrophotometric standards were taken from 
Hamuy et al. (\cite{Hamuy})
and the PN spectra were flux calibrated using iraf.noao.onedspec 
routines. Fig. \ref{MOSspec} shows an example of the reduced multi-slit
spectra for field F56. The unequal wavelength coverage of the 
spectra is dependent on the relative position of the target 
centred slitlet within the field. Thus a target to the right 
edge of the field is truncated to the red but extends to lower 
wavelengths than a target in the centre of the FORS1 MOS field.

Fig. \ref{MOSspec} demonstrates that there can be more than one PN per slitlet; 
in some cases this was planned, by placing the slitlet such that two 
PN from the Hui et al. catalogue lay on the slit, but in some cases 
in the two inner fields, PN were spectroscopically detected that were not 
present in the Hui et al. catalogue, on account of their faintness 
or proximity to another catalogued PN. 
These previously uncatalogued PN typically had m$_{5007A} > 27$mag. 
On the basis of the strongest line - [O~III]5007\AA\ - the number of 
PN detectable in the three fields were: 21 in F56; 21 in F42; and 9 
in F34. On each slitlet, regions distinct from the PN spectra were 
located and used to subtract the background, with a 1st or 2nd order 
fit in the cross-dispersion direction. In determining the spectra of 
the PN, no attempt was made to perform a separate sky background 
subtraction since there is galaxy background at all slitlet positions, 
except perhaps for the slitlets at largest galactocentric distance 
in the outermost field (F34). 1D spectra of each PN were
formed by summing the background subtracted signal along the slit; 
7-9 pixels (1.4-1.8$\arcsec$) were used to collect around 60 to 
70\% of the flux for 0.9$\arcsec$ seeing (assuming a Gaussian PSF).
In addition to producing 1D spectra of flux, the same extraction
was performed for non-flux calibrated spectra in order to determine
the random noise errors on the extracted spectra. No attempt was made 
to propagate other sources of error, such as flat fielding, into
the resultant error vectors 

As is evident from Tab. \ref{PNSpeObs} all fields were observed on more than 
one occasion. In order not to lose any information, all the flux
calibrated and extracted 1D spectra were employed to form averaged spectra
per PN. It was expected that the average should be formed using the 
inverse of the exposure time as weight; however
the sky transmission, airmass, seeing and moon phase
could all affect the resulting noise in a spectrum, so a simple unweighted
mean was finally employed. However the F56 600B spectra obtained in 
2003 were found to have lower fluxes but similar signal-to-noise (S/N) as the
2000 observations. These later observations were rescaled in forming the
average. For each PN, the various spectra were intercompared before 
averaging; if a line was found to be significantly discrepant between spectra, 
a careful examination of the raw 2D spectra was made to determine if a 
processing step, such as incorrect CR removal in the PN spectrum or the 
local background, had affected the emission line flux. If a line was deemed 
to be badly affected, the final spectrum was substituted using only the clean 
spectrum. 

The 300V spectra overlap with the 600B spectra for the range 
4600 - 5800\AA\ allowing the red lines (He~I 5876\AA, 
H$\alpha$+[N~II], [S~II], etc) to be placed on the same relative 
flux scale as the blue lines, primarily using the strong 5007\AA\ line 
to scale the spectra. The errors on the final spectra were combined using 
Gaussian error propagation though these combination steps. The 
resultant spectra thus have a range of line S/N and the magnitude of 
the propagated errors were checked in two ways: regions of line-free 
continua were chosen and the root-mean square on the mean was compared 
with the mean of the statistical errors for the same regions; the rms 
on the mean value of the fixed [O~III]5007/4959\AA\ ratio was compared
with the mean of the errors on this observed ratio from the Gaussian
line fits (see Section 4.1). It was found that the 
naively propagated errors over-estimated the real errors on the
data values as demonstrated by these two tests; the factor varied
slightly between data sets but was around 1.7 for each spectrum. The 
statistical errors in each spectrum were amended by this amount and 
are those that are listed as the propagated errors on the measured line 
fluxes (see Tab. \ref{ObsFlux}).

In order to examine the stellar spectra at the positions of the PN, 
an attempt was made to subtract the sky background for the bluer 
(600B) spectra. The F34 field contains the highest proportion of 
sky over galaxy background, so was used to remove sky from the 
exposures of the other fields when F34 and other fields were 
observed on the same night (observing runs in 2001 and 2003 - 
see Tab. \ref{PNSpeObs}). The flux calibrated sky spectrum per pixel for the 
F34 field, excluding the PN spectra, was formed and subtracted 
from the F56 and F42 data. Some mismatches in terms of the sky 
lines were noted, particularly [N~I],
and some of the sky spectra produced poor subtraction in the sense 
of over-subtraction to short wavelengths. These problematic spectra 
were not used in forming a mean galactic background spectrum in the 
two fields F42 and F56.

\section{Results}

\subsection{Individual PN spectra}
The 1D spectra of the 51 PN in NGC~5128 were analysed by interactively
fitting Gaussians to the emission lines with a linear interpolation to 
the underlying galaxy continuum over the line extent. Errors on the 
line fits were propagated from the flux errors. The extinction correction 
was determined by comparison with the case B values for 12000K
and 5000cm$^{-3}$ and the Seaton (\cite{Seaton}) Galactic reddening 
law with R=3.2 (in the absence of other information on the appropriate reddening 
law to adopt for NGC~5128), in all cases where at least the H$\alpha$ and 
H$\beta$ lines were detected. The observed line fluxes (normalised to 
H$\beta$=100) and errors are presented in Tab. \ref{ObsFlux}. The field 
number and slitlet numbering of the target is provided, the number 
from the catalogue of Hui et al (\cite{Hui93b}) together with the
$\Delta \alpha$, $\Delta \delta$ offsets (in arcsec) from the position 
of the nucleus (taken as the SIMBAD coordinate 
13$^{h}$ 25$^{m}$ 27.6$^{s}$ -43$^\circ$ 01$'$ 08.8$''$ (J2000)), 
the observed logarithmic H$\beta$ line flux and error, and the 
m$_{5007A}$ determined from the observed [O~III]5007\AA\ line flux
and error. The extinction correction c, and $E_{B-V}$ values
are listed in Tab. \ref{DeredFlux}. The errors on the extinction 
were not propagated to the dereddened line errors. 

Seven PNe with weak emission lines, whose large errors on the H$\beta$ 
line precluded reliable determination of line ratios,
were also analysed but are not included in Tab. \ref{ObsFlux};
these are listed separately in Tab. \ref{FaintPN}. Except for the
PN F56\#7 (5509 in Hui et al. \cite{Hui93b}), the absolute 
coordinates of these faint PN were determined from the relative 
positions of the PN on the slitlet with respect to the PN coordinates 
from Hui et al.(\cite{Hui93b}). The [O~III]/H$\beta$ ratios are listed 
in Tab. \ref{FaintPN} where H$\beta$ was detected; 
however given the large uncertainty on the H$\beta$ line flux, these 
ratios are poorly determined. The PN not detected from the narrow band 
imaging of Hui et al. (\cite{Hui93b}) are all faint (m$_{5007A}$ $>$ 26.5) 
and are near the centre of the galaxy where the stellar continuum 
is high. Slit spectroscopy allows lower equivalent width emission 
lines to be detected and can thus probe deeper than filter imaging. 

\begin{table*}
\caption[]{NGC~5128 PN without reliably measured H$\beta$ fluxes}
\label{FaintPN}
\centering 
\begin{tabular}{l l l c l l l} 
MOS Slitlet & Hui et al. & ~~~~~~~$\alpha$ & ~~~~~$\delta$ & ~~~~~Offsets & m$_{5007A}$ & [O~III]/H$\beta$ \\ 
            & No.        & ~~($h$ $m$ $s$) & ($^\circ$ $'$ $''$) & ~~~~~$\Delta \alpha$ $\Delta \delta$ & mag.$^\ast$ & approx. \\
            &      &             &               & ~~~~(arcsec) &                      &      \\
\hline
F56\#7      & 5509 & 13 25 56.41 & $-$43 06 23.2 & ($+$293.6,$-$314.4) & 26.2$^\dag$ & 29 \\
F56\#13a    &      & 13 25 46.50 & $-$43 04 57.3 & ($+$207.3,$-$228.5) & 27.6 & 4 \\
F56\#13c    &      & 13 25 45.78 & $-$43 04 53.3 & ($+$199.4,$-$224.5) & 27.5 & 13 \\
F42\#12a    &      & 13 25 15.21 & $-$43 03 12.5 & ($-$135.9,$-$123.7) & 26.5 & 23 \\
F42\#14a    &      & 13 25 18.20 & $-$43 02 23.1 & ($-$103.1,$-$74.3)  & 26.8 & 22 \\
F42\#14c    &      & 13 25 19.24 & $-$43 07 19.9 & ($-$91.7,$-$71.1)   & 27.1 & 9 \\
F42\#16b    &      & 13 25 23.13 & $-$43 02 49.6 & ($-$49.0,$-$100.8)  & 27.2 & 9 \\
\end{tabular}
\tablefoot{
$^\ast$ Spectroscopic magnitude, corrected for slit/pointing losses. \\
$^\dag$ Hui et al. \cite{Hui93b} list m$_{5007A}$ of 25.76.
}
\end{table*}

\begin{figure}
\centering
\resizebox{\hsize}{!}{\includegraphics[width=10.0cm]{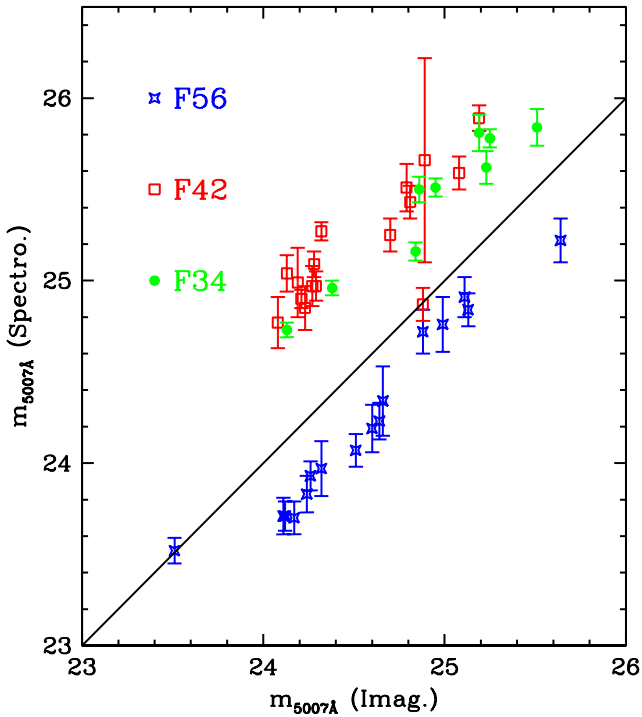}}
\caption{Comparison of the imaging m$_{5007A}$ magnitude from Hui 
et al. (\cite{Hui93b}) with the equivalent magnitude computed from the 
observed [O~III]5007\AA\ spectroscopic flux converted to
magnitude. The PN in the three fields are shown with different symbols,
as indicated.}
\label{O3mags}
\end{figure}

The comparison between the m$_{5007A}$ magnitudes from this work, 
converting the measured 5007\AA\ flux to m$_{5007A}$, and those from Hui et al. 
(\cite{Hui93b}) is shown in Fig. \ref{O3mags}. For Field 
F56, the observed m$_{5007A}$ magnitudes are on average 0.33$\pm$0.12 mag. 
brighter than the Hui et al. measurements, excluding the two faintest PN. 
The absolute H$\beta$ flux calibration was adopted from
the May 2000 600B observations; presumably the spectrophotometric 
standard was taken under slightly poorer conditions, leading to an 
apparent higher absolute flux for the PN observations.
For the F42 field, the m$_{5007A}$ magnitudes are 0.67$\pm$0.21 mag. fainter 
than the Hui et al. ones, which can be attributed to an offset of 
the slits from the PN positions, since all the
F42 observations display lower m$_{5007A}$ than the imaging 
observations. For the F34 field, the m$_{5007A}$ magnitudes are on 
average 0.51$\pm$0.21 mag. fainter than the imaging magnitudes, which
are consistent with a $\approx$0.2$''$ slit offset for 0.8$''$ wide 
slits in 1.0$''$ seeing (assuming a Gaussian seeing profile).

\subsection{Region-averaged PN spectra}
Single spectra consisting of the sum of all the observed brighter PN 
(i.e. those contained in Tab. \ref{ObsFlux}) in each of the
three fields were constructed. 
%ADD 
The primary motivation was to enable the detection of fainter 
diagnostic lines (such as He~I, [S~II], [Ar~IV], etc.) which were not
detected, or were marginally detected, on the individual spectra. This
was also an explorative study to measure how different the
derived abundances would be from the summed spectra in comparison with
the ensemble of the individual PN abundances. This aspect is of 
interest for studies of more distant PN populations where all line 
detections from individual PN spectra have low S/N and summed spectra 
are mandatory for measurement of abundances.
The radial velocity of each PN was measured 
from the strong lines in the spectrum (principally [O~III] 4959 and 5007\AA\
and H$\alpha$) and employed to shift all the spectra in a single region 
to zero radial velocity. The spectra were then summed and the resulting
emission line spectra fitted in the same way as the individual PN
spectra (Section 4.1). Figure 2 of Walsh et al. (\cite{Walsh05}) shows the 
resulting three spectra. Tab. \ref{RegObsFlux} lists the resulting observed 
fluxes. The extinction was calculated from the H line ratios with the same 
assumptions as for the individual PN spectra (Sect. 4.1) and the dereddened 
line fluxes are listed in Tab. \ref{RegDerFlux}.

The ionic abundances of the elements corresponding to the detected species can 
be determined from the dereddened line ratios with respect to H$\beta$ (Tab. 
\ref{RegDerFlux}). The electron temperature of the O$^{++}$ emitting region can 
be directly measured from the 5007/4363\AA, but also required for
abundances is an electron density measure. The only directly determined N$_e$ 
value is from the [S~II]6716/6731\AA\ ratio, which has two disadvantages in that it samples
the minority low ionization nebular volume and has a relatively low critical
density for collisional de-excitation. Coppeti \& Writzel (\cite{CoWR02}) show 
that [S~II] N$_e$ measured in PN is similar to N$_e$ measured for higher 
ionization species, such as [Cl~III] and [Ar~IV]. Adopting a value
of N$_e$ of 5000 cm$^{-3}$ for the NGC~5128 PN appears to be fair
and will not lead to bias on the abundance estimates, unless the density is very
high ($^{>}_\sim$20\,000 cm$^{-3}$). In the high density case, the 
[O~III]5007/4363\AA\ line ratio is decreased by collisional de-excitation, 
leading to too high a value of T$_e$.

Tab. \ref{Regabunds} lists the derived ionic abundances for the 
summed region spectra. The sources for the atomic data for the collisionally 
excited lines were taken from the compilation in Liu et al. (\cite{Liu2000})
and the routines from Storey \& Hummer (\cite{StorHum}) were used for the 
recombination line emissivities of H and He. The corrections for unseen stages 
of ionization (ionization correction factors, ICFs) were taken from 
Kingsburgh \& Barlow (\cite{KiBa}). The errors in the total abundances do not 
take into account the errors in the electron temperature; if these are 
propagated the errors rise to $\pm$0.10 for the O abundance.
The He/H abundance is only listed if the lines at 4471 and or 5876\AA\
were detected; if He~II 4686\AA\ was detected in addition, it was added
to derive the total He/H. If only He II was detected no He/H is listed.

\begin{table*}
\caption[]{Summed PN spectra of three fields in NGC~5128 - Observed line fluxes}
\label{RegObsFlux}
\centering 
\begin{tabular}{l l | r r | r r | r r }
         &                   & \multicolumn{2}{c}{F56} & \multicolumn{2}{c}{F42} &  \multicolumn{2}{c}{F34} \\
\hline
 Species & $\lambda$ (\AA)  & F$_{Obs}$ & $\pm$ &  F$_{Obs}$ & $\pm$ & F$_{Obs}$ & $\pm$ \\
\hline
~[O~II]                 & 3727 &   28.6 &  1.7 &   23.9 &  2.2 &   54.5 &  2.9  \\
~[Ne III]               & 3868 &   84.0 &  1.7 &   62.3 &  1.5 &   74.0 &  1.9  \\
~H~I                    & 3889 &   17.3 &  4.4 &   13.3 &  1.6 &   14.3 &  1.2  \\
~[Ne~III] + H$\epsilon$ & 3970 &   33.5 &  1.0 &   26.9 &  1.5 &   34.4 &  1.9  \\
~H$\delta$              & 4101 &   21.1 &  1.2 &   10.7 &  1.2 &   21.3 &  1.1  \\
~H$\gamma$              & 4340 &   46.7 &  1.9 &   41.5 &  2.0 &   40.8 &  1.5  \\
~[O~III]                & 4363 &    8.7 &  0.7 &    6.1 &  1.3 &    8.5 &  0.6  \\
~He~I                   & 4471 &    8.0 &  1.0 &    6.5 &  1.0 &    5.0 &  3.3  \\
~He~II                  & 4686 &   10.7 &  1.2 &    9.3 &  1.9 &   13.1 &  1.1  \\
~H$\beta$               & 4861 &  100.0 &  0.0 &  100.0 &  0.0 &  100.0 &  0.0  \\
~[O~III]                & 4959 &  391.3 &  5.7 &  396.4 &  7.1 &  417.1 &  7.1  \\
~[O~III]                & 5007 & 1159.6 & 16.0 & 1213.6 & 20.2 & 1247.8 & 20.6  \\
~He~I                   & 5016 &    3.7 &  0.4 &        &      &        &       \\
~[N~I]                  & 5199 &    3.2 &  0.2 &        &      &        &       \\
~He~I                   & 5876 &   14.5 &  1.9 &   29.9 &  5.6 &        &       \\
~[N~II]                 & 6548 &   50.3 &  1.2 &   62.7 &  4.4 &   62.7 &  4.6  \\
~H$\alpha$              & 6562 &  390.6 &  3.8 &  422.7 &  9.8 &  412.8 & 14.7  \\
~[N~II]                 & 6583 &  152.2 &  1.8 &  149.2 &  4.3 &  115.7 &  5.5  \\
~He~I                   & 6678 &    4.6 &  0.8 &    7.1 &  4.6 &        &       \\
~[S~II]                 & 6716 &    4.5 &  0.4 &    7.3 &  2.8 &        &       \\
~[S~II]                 & 6730 &   10.9 &  1.1 &   27.9 &  3.0 &        &       \\
~[Ar~III]               & 7133 &   28.5 &  1.1 &   22.9 &  3.1 &        &       \\
~[O~II]                 & 7325 &   27.2 &  2.4 &   19.6 &  4.6 &        &       \\
\hline                
log F(H$\beta$)         &      & -15.46 & 0.01 &  -15.42 & 0.01 &  -15.78 & 0.01  \\
m$_{5007A}$             &      &  22.24 & 0.03 &   22.10 & 0.04 &   22.98 & 0.04  \\
\hline
\end{tabular}
\end{table*}

\begin{table*}
\caption[]{NGC~5128 PN Region spectra - Dereddened line fluxes}
\label{RegDerFlux}
\centering 
\begin{tabular}{l l | r r | r r | r r }
         &                   & \multicolumn{2}{c}{F56} & \multicolumn{2}{c}{F42} &  \multicolumn{2}{c}{F34} \\
\hline
 Species & $\lambda$ (\AA)  & F$_{Obs}$ & $\pm$ &  F$_{Obs}$ & $\pm$ & F$_{Obs}$ & $\pm$ \\
\hline
~[O~II]                 & 3727 &   37.0 &  2.3 &   33.0 &  3.0 &   73.8 &  3.9  \\
~[Ne III]               & 3868 &  105.9 &  2.1 &   83.3 &  2.1 &   97.2 &  2.5  \\
~H~I                    & 3889 &   72.0 &  5.5 &   17.7 &  2.1 &   18.7 &  1.5  \\
~[Ne~III] + H$\epsilon$ & 3970 &   41.4 &  1.2 &   35.0 &  1.9 &   44.2 &  2.4  \\
~H$\delta$              & 4101 &   25.4 &  1.4 &   13.5 &  1.4 &   26.4 &  1.4  \\
~H$\gamma$              & 4340 &   53.1 &  1.8 &   48.7 &  2.4 &   47.4 &  1.7  \\
~[O~III]                & 4363 &    9.8 &  0.8 &    6.6 &  1.3 &    9.9 &  0.6  \\
~He~I                   & 4471 &    8.9 &  1.0 &    7.4 &  1.2 &    5.6 &  3.6  \\
~He~II                  & 4686 &   11.2 &  1.3 &    9.8 &  2.0 &   13.8 &  1.1  \\
~H$\beta$               & 4861 &  100.0 &  0.0 &  100.0 &  0.0 &  100.0 &  0.0  \\
~[O~III]                & 4959 &  381.8 &  5.6 &  384.6 &  6.9 &  405.5 &  6.9  \\
~[O~III]                & 5007 & 1118.3 & 15.4 & 1160.0 & 20.0 & 1196.3 & 19.7  \\
~He~I                   & 5016 &    3.6 &  0.4 &        &      &        &       \\
~[N~I]                  & 5199 &    3.0 &  0.2 &        &      &        &       \\
~He~I                   & 5876 &   11.6 &  1.5 &   22.8 &  4.3 &        &       \\
~[N~II]                 & 6548 &   36.5 &  0.9 &   42.1 &  2.9 &   43.0 &  3.1  \\
~H$\alpha$              & 6562 &  282.9 &  2.8 &  282.8 &  6.6 &  282.8 & 10.1  \\
~[N~II]                 & 6583 &  109.9 &  1.3 &   99.4 &  2.9 &   79.0 &  3.7  \\
~He~I                   & 6678 &    3.3 &  0.5 &    4.7 &  3.0 &        &       \\
~[S~II]                 & 6716 &    3.2 &  0.3 &    4.8 &  1.8 &        &       \\
~[S~II]                 & 6730 &    7.7 &  0.8 &   18.1 &  1.9 &        &       \\
~[Ar~III]               & 7133 &   19.1 &  0.8 &   13.9 &  1.9 &        &       \\
~[O~II]                 & 7325 &   17.9 &  1.6 &   11.6 &  2.7 &        &       \\
\hline                
c                       &      &  0.44  & 0.03 &    0.55 & 0.07 &    0.51 & 0.12  \\
log F(H$\beta$)         &      & -15.03 & 0.01 &  -14.88 & 0.01 &  -15.27 & 0.01  \\
m$_{5007A}$             &      &  21.18 & 0.03 &   20.74 & 0.04 &   21.74 & 0.04  \\
\hline
\end{tabular}
\end{table*}

\begin{table*}
\caption[]{NGC~5128 PN Region spectra - Diagnostics and Abundances}
\label{Regabunds}
\centering 
\begin{tabular}{l | r r | r r | r r }
 Species & \multicolumn{2}{c}{F56} & \multicolumn{2}{c}{F42} &  \multicolumn{2}{c}{F34} \\
\hline
         & Value & $\pm$ & Value & $\pm$ & Value & $\pm$ \\
\hline
~[O~III] (5007+4959)/4363\AA & 153.1 & 12.6 &  234.0 &  46.2 & 161.8 & 10.0 \\
~[O~III] T$_e$ (K)           & 10970 & 300  &   9620 &   560 & 10780 &  220 \\
                             &       &      &        &       &       &       \\
~O$^{+}$   & 1.76E-5 & 1.1E-6 & 2.69E-5 & 2.4E-6 & 3.52e-5 & 1.8E-6 \\
~Ne$^{++}$ & 7.93E-5 & 1.6E-6 & 1.03E-4 & 2.5E-6 & 7.28E-5 & 1.9E-6 \\
~He$^{+}$  & 0.173 & 0.019 & 0.142 & 0.023 & 0.109 & 0.070 \\
~He$^{++}$ (4471\AA) & 0.0094 & 0.0011 & 0.0081 & 0.0016 & 0.0115 & 0.0009 \\
~O$^{++}$  & 3.091E-4 & 4.3E-6 & 4.997E-4 & 8.6E-6 & 3.307E-4 & 5.4E-6 \\
~He$^{+}$ (5876\AA)  & 0.084 & 0.011 & 0.160 & 0.031 &       &       \\
~N$^{+}$   & 1.83E-5 & 2.2E-7 & 2.28E-5 & 6.6E-7 & 1.32E-5 & 6.2E-7 \\
~S$^{+}$   & 4.40E-7 & 4.6E-8 & 1.39E-6 & 1.5E-7 &        &       \\
~Ar$^{++}$ & 1.50E-6 & 6.3E-8 & 1.46E-6 & 2.0E-7 &        &       \\
\hline                
12 + Log(O/H) & 8.53 & 0.01 & 8.74 & 0.01 & 8.59 & 0.02  \\
He/H          & 0.182 & 0.019 & 0.150 & 0.023 & 0.122 & 0.0070 \\
Ne/O          & 0.257 & 0.006 & 0.206 & 0.006 & 0.220 & 0.007 \\
N/O           & 1.04  & 0.07  & 0.85  & 0.08  & 0.38 & 0.03 \\
12 + log(S/H) & 6.55  & 0.05  & 7.06  & 0.04  &      &      \\ 
12 + log (Ar/H) & 6.49 & 0.04 & 6.44  & 0.06  &      &     \\ 
%m$_{5007A}$     &  22.24 & 0.03 & 22.10 & 0.04 & 22.98 & 0.04  \\
\hline
\end{tabular}
\end{table*}

\subsection{Stellar absorption lines}
\begin{figure}
\centering
\resizebox{\hsize}{!}{\includegraphics[width=10.0cm,angle=0]{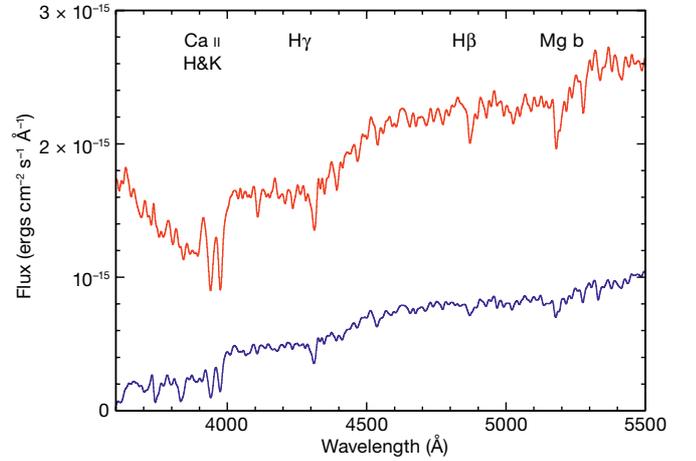}}
\caption{The observed spectra of the galaxy stellar continuum, summed along
the whole slit but excluding the positions occupied by the PN, are shown for 
fields F42 (upper spectrum, in red) and F56 (lower spectrum,in blue). The 
strongest absorption lines are identified.}
\label{Galcont}
\end{figure}

The slitlets sample a spectroscopic background consisting of sky with a 
substantial contribution from the galaxy continuum, except for the outermost 
field, F34. The regions of the slitlets not occupied by 
planetary nebula spectra can provide the spectrum of the stellar continuum 
of NGC~5128. 
%ADD
The inner fields, F56 and F42, contain no slitlets sampling the sky 
free of stellar continuum, and so no simultaneous sky subtraction can be 
performed. For the outer field F34 however, the stellar continuum is 
very weak and was not detected, as shown by the absence of a gradient in 
the background with radial offset from the galaxy centre.
The 600B F34 spectrum taken on 2003-04-08 was thus used to produce a 
candidate mean sky spectrum for subtraction from the spectra for the 
two inner fields (F56 and F42). 
This sky was interactively subtracted from the F42 and F56 spectra for 
each slitlet. Good results were found only for the F42 spectra
taken in April 2001 and for the F56 field spectra taken in May 2003. These
results are attributable to very differing sky contributions to the galaxy spectra
at the different observation epochs. For the F56 field, the sky-subtracted 
data show very low continuum in the five slitlets (1-5) furthest from the 
galaxy centre; these slitlets were not considered for analysis of the 
galaxian stellar continuum. Fig. \ref{Galcont} shows the resulting stellar 
continuum from fields F56 (blue line) and F42 (red line), for all
the slitlets summed. Strong CaII H \& K, H$\gamma$, H$\beta$ and Mg II lines 
typical of an intermediate-old stellar population are clearly visible
and are indicated on the Figure.

\begin{figure}
\centering
\resizebox{\hsize}{!}{\includegraphics[width=10.0cm,angle=-90,bb=22 100 596 720,clip]{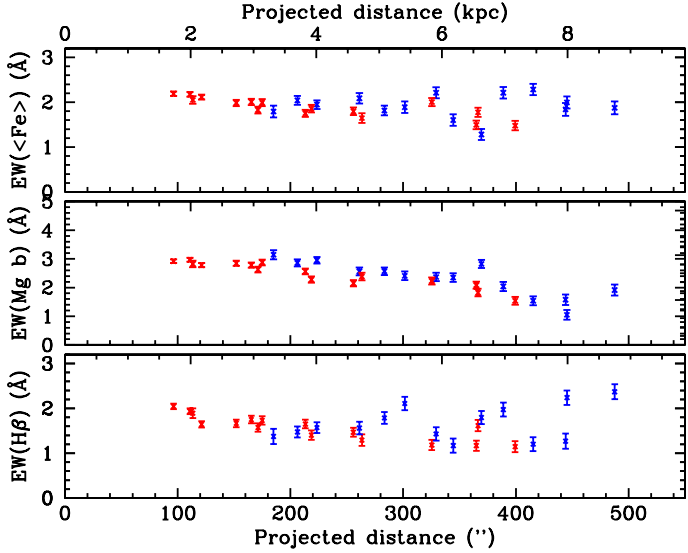}}
% Bounding box of figure = 22 16 596 784
% trim=400 100 2500 250
\caption{The Lick indices, EW(H$\beta$), EW(Mgb) and 
$<$Fe$>$ (defined as 0.5*(EW(5270\AA) + EW(5335\AA)), determined from the 
stellar continuum in each slitlet, are plotted as a function of the projected
distance from the nucleus of NGC~5128, in arcsec (lower axis) and kpc (upper 
axis). Measurements for field F42 are shown in red and for field F56 in blue.}
\label{Licks}
\end{figure}

  The absorption lines in the individual galaxy continuum spectra for each 
slitlet were analysed by computing Lick indices (Worthey et al. \cite{Worthey}).
The spectra were smoothed with a Gaussian to simulate the $\sim$8\AA\ resolution
of the Lick/IDS stellar spectra and shifted to zero velocity before the
equivalent widths were computed. Errors on the indices were computed by 
propagating the flux errors in the equivalent width determination. A single
radial velocity of 580 kms$^{-1}$ was used to shift the spectra to rest
wavelength. Errors of $\pm$50 kms$^{-1}$ in this radial velocity cause an
offset in the bounds of the Lick indices and can introduce errors
up to about 0.2\AA\ in EW. Three of the Lick indices (H$\beta$, Mg\,b and 
the combined Fe index $<$Fe$>$, defined as 0.5*[EW(Fe$_{5270A}$) + 
EW(Fe$_{5335A}$)], are plotted in Fig. \ref{Licks} as a function 
of the projected position of the slitlet centres from the centre of the
galaxy. The $<$Fe$>$ index shows no radial gradient nor does the 
H$\beta$ index, although the latter shows higher scatter to larger
projected distance from the nucleus. The Mg b index shows a
weak negative radial gradient, typical of early type galaxies (c.f. Davies 
et al. \cite{Davies93}).

\section{Chemical abundance determination}

\subsection{Extragalactic PN abundances}

The electron temperature (T$_e$) is required In order to determine reliable 
chemical abundances for extra-galactic PNe from the collisionally excited 
lines on account of the exponential dependence of 
line emissivity with temperature. The electron density (N$_e$) is also required 
since these lines have collision cross sections dependent on density. 
In order to avoid large corrections for unseen ionization species a range of 
lines of different ionization from neutral and singly ionized up to high 
ionization is also desirable; however the well-established empirical technique 
using ICFs can compensate for the lack of some ionization species from the 
optical wavelength range. As is typical for faint and distant PNe
such as observed here, these criteria are not fully met. In the absence of 
measurements of weak T$_e$ and N$_e$ diagnostic line ratios (such as [O~III]
5007/4363\AA\ and [S~II]6716/6731\AA\ or [Ar~IV]4711/4740\AA), the practice 
adopted by Stasi\'{n}ska et al. (\cite{SRM98}) was to use the upper limits to the 
strength of these faint lines to constrain T$_e$ and N$_e$ within reasonable 
limits, taking typical values for PN 
from studies in the Milky Way and Magellanic Clouds (e.g. T$_e$ $\sim$12000K, 
N$_e$ $\sim$5000 cm$^{-3}$). Jacoby \& Ciardullo (\cite{JC99}) introduced a different 
approach to abundance determination for extra-galactic PNe (in this
case for M31): that of photoionization modelling using the very well-established
Cloudy (Ferland et al. \cite{Ferland}) code. 
The gains of this latter method are that both strong lines with low errors, 
weak lines with large errors and upper limits can all be used 
in arriving at a satisfactory model, {\em and} the parameters of the central stars 
(luminosity, $\cal L$ and effective temperature, $T_{eff}$) are derived as part of the 
photoionization modelling. 

Extra-galactic PNe are spatially unresolved but have known distances 
and absolute fluxes available though m$_{5007\AA}$ photometry. An input 
data set for photoionization modelling can be assembled, given some simplifying 
assumptions in the absence of more detailed information; in particular, the 
nebular geometry and information on the stellar atmosphere are not available. 
A black body has been shown to be an acceptable assumption for PN central stars 
(Howard et al. \cite{Howard97}), which leaves the 
nebular geometry. Taking a spherical shell of constant density as a baseline, 
Jacoby \& Ciardullo (\cite{JC99}), hereafter JC99, compared nebular abundances 
using directly measured T$_e$ and N$_e$ values and ICFs to check that 
photoionization modelling was indeed a valid approach. The approach was
shown to work well. Additional complexity, such as two zone density structure 
and the presence of dust within the nebula, was found to be 
required to satisfactorily match the observed spectra in some cases. 

Magrini et al. (\cite{Magrini}) followed
the same procedure as JC99 in determining abundances of PNe in M33. They also
tested the method by applying it to the integrated spectra of well-observed Milky
Way PNe and compared the model abundances to those from the ICF method, finding
agreement within 0.15 dex for O/H and 0.3 dex for N/H. In addition Magrini et al. 
(\cite{Magrini}) compared abundances derived from Cloudy models with those from ICFs 
for three PNe of their M33 sample. They also considered comparisons employing 
model atmospheres and an $r^{-2}$ density in the nebular
shell, and found no major discrepancies. The technique, involving
the development of a large number of models in working to a satisfactory
match of observed and model spectrum, is labour-intensive and therefore 
not suitable for large samples. It also requires simplifying assumptions. 
The technique, however, is well suited to the present sample of
PN spectra in NGC 5128 where the range of abundances is of 
particular interest and the data set is limited; it also serves as a 
probe of the star formation history. 

Only 10 of the PN spectra presented in Tables \ref{ObsFlux} and \ref{DeredFlux} 
have detectable [O~III]4363\AA, enabling direct measurement of 
the electron temperature, and only 7 have an electron density sensitive 
ratio ([S~II]6716/6731\AA) measured. The typical S/N on the weak 
4363\AA\ line ($\leq 3$) implies that T$_e$ errors of
1250K at 11000K affect abundance determination, leading to errors of 
$^{+0.17}_{-0.13}$ on log (O$^{++}$/H$^{+}$). For the PNe without detectable 
diagnostic line ratios, two approaches are possible: employ the region-averaged
spectra to assign T$_e$ and N$_e$ values to the single PN, although these values 
may be substantially in error for particular nebulae; or, employ photoionization 
modelling of the strong lines to find the best-matching model that does not
violate the upper limit constraints for the weak lines.

\subsection{Modelling procedure}
The simplest possible photoionization model was employed: a black body central 
star characterized by its temperature and luminosity, the latter given by
the H$\beta$ luminosity derived from the m$_{5007\AA}$ magnitude (Hui et al.
\cite{Hui93b}), the reddening and the dereddened 5007\AA/H$\beta$; a spherical 
shell with the inner radius set to a small value (0.005 pc) to 
ensure high ionization emission close to the central star, but not 
so close that instabilities arise in the model; the outer radius set to 
a large value (0.5 pc) since the nebulae are expected to be optically thick, 
as given the high luminosity of the PNe observed in NGC~5128 (i.e. within 2 mag. of the
peak of the PNLF); and an initial set of abundances. For the abundances
of He, N, O, Ne, S and Ar, the initial values were taken from the ICF analysis 
for the integrated spectra listed in Tab. \ref{Cloumods} and the C/O ratio taken 
as 0.5. However these values were immediately allowed to vary to fit the 
individual spectra (with C varying in lock-step with O) and should not be seen
as prejudicing the individual determinations from the higher S/N integrated
spectra. The initial assumption of an optically thick nebula could also have 
been relaxed in the modelling process but was not found to be demanded by the 
fits to the PN spectra. The philosophy was to
aim to match the dereddened lines fluxes within the listed errors for all the 
spectra listed in Tab. \ref{DeredFlux} (except F56\#9, F56\#17, F42\#15b and
F42\#17 without detected [O~II] or He lines), thus 40 PN spectra. These PNe
represent the upper 2.1 mag. of the PNLF.

  The modelling procedure followed closely that outlined by JC99 and also Marigo
et al. \cite{Marigo}. An initial default density of 5000cm$^{-3}$ was used. The
initial estimate of $T_{eff}$ was taken from He~II 4686\AA/H$\beta$; in the absence of
4686\AA, its upper limit and the 5007\AA/H$\beta$ ratio was used. The presence
of 4686\AA\ enabled $T_{eff}$ to be refined within a few runs of Cloudy; if 4686\AA\
was not present, both $T_{eff}$ and O/H had to be adjusted. Then He was adjusted to
match the He I lines, with minor adjustments to T$_e$ for the concomitant changes in
He~II flux. O/H and the density were adjusted to match [O~II]3727\AA\ and, if 
present, the [O~II]7325\AA\ quadruplet. For large changes in O abundance, 
Ne and C were initially changed in proportion. Once the basic physical
parameters ($T_{eff}$, density, He, O) were roughly determined, Ne, N, Ar and S were
altered to match the observed lines. In the final phase all lines and  $T_{eff}$
and density were subject to small changes to improve the fit, where possible. A
guide to the goodness of fit of the observed and predicted line strengths was formed, 
taking the sum of the line flux differences normalised by the S/N: thus larger
discrepancies for weak lines could be weighted lower. A discussion of the goodness
of the matches and the estimation of errors in presented in section 5.4.
In general around 10 iterations per PN was enough to reach a satisfactory match to 
the spectrum but in more difficult cases (F56\#12b, F42\#10 and F34\#7 for example)
up to 30 iterations were required. In total, $\sim$600 Cloudy models were run  to 
complete this analysis.

\begin{table*}
\caption[]{Cloudy photoionization modelling of 40 PN in NGC~5128}
\label{Cloumods}
\centering 
\begin{tabular}{l | r r r r r r | r c r c} 
\hline\hline
PN \#  & \multicolumn{6}{c}{Elemental Abundances 12 + log(A/H)} & T$_{eff}$ & Log $L$ & Ne~~~ & Quality$^{3}$ \\
       & (He/H) & (N/H) & (O/H) & (Ne/H) & (S/H)$^1$ & (Ar/H)$^2$ & (kK) & ($L_{\odot}$) & (cm$^{-3}$) & \\
\hline
F56\#1 & 11.06 & 8.20 & 8.40 & 7.70 & 6.55 & 6.00 & 60 & 3.68 & 5000 & b \\
F56\#2 & 11.15 & 7.85 & 8.43 & 7.72 &      & 6.00 & 155 & 4.20 & 10000 & a \\
% Revised to f56-1_604.in
F56\#3 & 11.06 & 8.30 & 8.48 & 7.78 & 6.55 & 6.00 & 64 & 3.66 & 2000 & b \\
F56\#4 & 11.06 & 8.30 & 8.38 & 7.70 & 6.55 &  6.50 & 78 & 3.75 & 5000 & b \\
F56\#5 & 11.06 & 7.40 & 8.55 & 7.85 &   &  6.35 & 150 & 3.74 & 10000 & c \\
F56\#6 & 11.20 & 8.30 & 8.64 & 7.95 &   &  6.60 & 75 & 4.15 & 6000 & b \\
F56\#8 & 11.20 & 8.62 & 8.60 & 7.80 & 6.80 & 6.50 & 63 & 3.77 & 9000 & b \\
F56\#10 & 11.10 & 8.20 & 8.38 & 7.60 & 6.55 &  6.60 & 88 & 4.03 & 15000 & b \\
F56\#11 & 10.90 & 8.42 & 8.54 & 7.70 & 7.00 & 6.40 & 90 & 3.67 & 22500 & a \\
F56\#12a & 11.10 & 8.53 & 8.80 & 8.10 & 6.55 &  6.40 & 200 & 3.86 & 10000 & a \\
F56\#12b & 11.00 & 8.80 & 8.55 & 7.85 &   &  6.50 & 45 & 3.93 & 35000 & b \\
F56\#13b & 11.00 & 7.60 & 8.52 & 7.70 &   &  6.20 & 105 & 3.76 & 20000 & a \\
F56\#14 & 11.06 & 8.40 & 8.46 & 7.70 & 6.55 &  6.30 & 95 & 3.86 & 15000 & b \\
F56\#15 & 11.06 & 7.90 & 8.47 & 7.70 & 6.55 &  6.20 & 70 & 3.86 & 10000 & c \\
F56\#16 & 11.06 & 8.20 & 8.47 & 7.60 & 6.55 &  6.20 & 70 & 3.91 & 10000 & b \\
% F56\#17 & 11.10 & 7.90 & 8.45 & 7.40 &   &   & 60 & 3.70 & 10000 & c \\
F56\#18 & 11.06 & 8.63 & 8.51 & 7.85 & 6.90 &  & 90 & 3.87 & 15000 & b \\
       &       &      &      &      &     &   &    &       &       \\
F42\#1 & 11.12 & 7.85 & 8.45 & 7.65 & 6.90 & 6.20 & 80 & 3.83 & 10000 & c \\
F42\#2 & 11.06 & 8.19 & 8.68 & 8.03 &   &   & 155 & 4.05 & 10000 & a \\
F42\#3 & 11.06 & 8.26 & 8.44 & 7.50 &   &  6.40 & 50 & 3.99 & 10000 & c \\
% F42\#4 new values 02/09/11
F42\#4 & 11.01 & 8.12 & 8.03 & 7.10 & 6.40 & 6.20 & 86 & 4.00 & 12000 & a \\
F42\#6 & 11.15 & 7.30 & 8.44 & 7.60 & 7.10 &  & 63 & 3.63 & 13000 & b \\
F42\#7 & 11.06 & 8.00 & 8.51 & 7.70 &   &   & 127 & 3.64 & 10000 & b \\
F42\#8 & 11.06 & 8.20 & 8.46 & 7.70 &   &   & 93 & 3.86 & 15000 & b \\
F42\#9 & 11.12 & 8.20 & 8.45 & 7.70 &   &  6.50 & 74 & 3.78 & 30000 & b \\
F42\#10 & 11.28 & 8.18 & 8.10 & 7.40 & 6.85 & 6.10 & 120 & 4.29 & 2500 & a \\
F42\#11 & 11.06 & 8.25 & 8.28 & 7.35 & 6.85 & 6.20 & 100 & 3.72 & 10000 & c \\
F42\#12b & 11.06 & 8.22 & 8.69 & 8.07 &   &   & 108 & 4.09 & 10000 & b \\
F42\#13 & 11.06 & 8.28 & 8.90 & 8.25 &   &   & 190 & 3.84 & 16000 & a \\
F42\#14b & 11.06 & 8.20 & 9.00 & 8.40 & 6.70 &  & 220 & 4.06 & 20000 & a \\
F42\#16a & 11.06 & 8.28 & 8.72 & 7.92 &   &   & 100 & 3.78 & 14000 & a \\
% F42\#17 & 11.06 & 8.22 & 8.49 & 7.78 &   &   & 49 & 4.23 & 10000 & c \\
F42-18 & 11.06 & 7.68 & 8.49 & 7.45 &   &   & 120 & 4.35 & 10000 & c \\
       &       &      &      &      &     &   &    &       &       \\
F34\#1 & 11.06 & 8.25 & 8.73 & 8.00 &   &  5.65 & 71 & 3.89 & 8500 & a \\
F34\#2 & 11.23 & 7.62 & 8.09 & 7.38 &   &   & 90 & 4.11 & 1500 & a \\
F34\#4 & 11.06 & 8.10 & 8.46 & 7.70 &   &  6.30 & 67 & 3.90 & 8000 & b \\
F34\#7 & 11.06 & 8.30 & 8.91 & 8.30 &   &  6.50 & 82 & 3.99 & 13000 & a \\
F34\#11 & 11.16 & 7.40 & 8.53 & 7.78 &   &   & 180 & 3.59 & 30000 & a \\
F34\#12 & 11.16 & 7.35 & 8.48 & 7.30 &   &   & 167 & 3.13 & 30000 & a \\
F34\#14 & 11.06 & 7.90 & 8.62 & 7.82 &   &   & 70 & 3.61 & 20000 & b \\
F34\#15 & 11.06 & 7.40 & 8.43 & 7.76 &   &   & 123 & 3.90 & 10000 & b \\
F34\#16 & 11.06 & 8.28 & 8.63 & 7.95 & 6.55 &   & 180 & 3.70 & 5000 & a \\
\hline
\end{tabular}
\tablefoot{
$1$ 12 + log(S/H) listed if default value of 6.55 not applied. \\
$2$ 12 + log(Ar/H) listed if default value of 6.00 not applied. \\
$3$ Quality. a: He~II and [O~II] both detected; b: [O~II] but not He~II 
detected, or He~II detected but not [O~II]; c: neither He~II nor [O~II] detected.
}
\end{table*}

\subsection{Results and scrutiny} 
Tab. \ref{Cloumods} lists the derived abundances for He, N, O, Ne, Ar, and 
S with respect 
to H where lines of these elements were detected, together with 
the stellar luminosity, black body temperature and the constant shell density. The
quality of the derived parameters are broadly classified into three categories 
based on: He~II (and or [O~III]4363\AA) and [O~II] both detected (15 PN: grade a); 
[O~II] but not He~II detected (18 PN: grade b); 
neither He~II nor [O~II], but at least He~I, [O~III]5007\AA, [Ne~III]3868\AA\ and
[N~II]6583\AA, detected (6 PN: grade c). Ten of the quality ``a" spectra had 
4363\AA\ detected and the models tailored to fit the strong lines (i.e. a model 
was not tweaked to exactly fit the 4363\AA\ line) were found to match the 4363\AA\
line within the $1\sigma$ errors in all cases except one (F42\#16a; 4207 in Hui et 
al. \cite{Hui93b}). The results presented are not claimed to be unique, as the 
adopted geometry of the nebulae is the simplest possible and the use of a black body 
atmosphere does not produce the most satisfactory fits to the level of ionization
in detailed photoionization modelling of Galactic PNe (e.g. Wright et al.
\cite{Wright}). However these simple models are capable of matching the observed 
variety of spectra with a very plausible range of abundances, central star 
temperatures and nebular densities.

Carbon is the most important coolant after O, but no lines of C were 
detected so there are no direct constraints on the C/H abundance. In the 
absence of other evidence, the abundance of C was assumed 
to be similar to O (C/O=0.5) and to vary in lock-step with O as the 
abundance of O was altered to match the spectrum. Once an adequate model had
been found, tests were performed on four of the spectra, 
altering the C abundance
by a factor 2.5 and refitting for the other species. A maximum difference
in 12+log(O/H) of -0.10 was found for 12+log(C/H) higher by +0.40; for 
12+log(C/H) lower by 0.40, a maximum difference of +0.02 in 12+log(O/H)
was found. Smaller changes in N and Ne abundances were required to match
the spectra with these altered C abundances. Thus the abundances are not
very sensitive to modest changes in the assumed C/O ratio. 

For the best observed PN spectrum (viz. the target with the brightest 
m$_{5007\AA}$) PN, (F56\#2; 5601 in Hui et al. \cite{Hui93b}), a set of
Cloudy models were run with NLTE model atmospheres from the T\"{u}bingen compilation 
(Rauch \cite{Rauch03}), and also with the presence of dust inside the ionized region. 
H, He and PG1159 model atmospheres were employed. While the H 
model atmosphere showed a much lower $T_{eff}$, an He or PG1159 atmosphere 
at 170-180 kK produced satisfactory fits, with the resulting O/H lower by 
up to 0.16 dex. Including dust with an atmosphere model required that 
O/H be reduced by up to 0.05 for a gas-to-dust mass ratio of 1000;
as the primary coolant of the nebula, the lower O/H is compensated by 
the enhanced cooling from dust.

Stasi\'{n}ska (\cite{Stas2002}) questioned the uniqueness of abundance determinations
even in the case that a direct measure of T$_e$ from e.g. [O~III]4363\AA\ is 
available. Depending on the temperature of the central star, a strong
gradient in the electron temperature can exist within the nebula and a single 
value of T$_e$ may not be a good representation for the bulk of the emission. 
The T$_e$ gradient is far less steep for PNe with hot central stars 
($\sim$100\,000K) than for cooler central stars and the use of a single T$_e$ 
should yield reliable abundances. Stasi\'{n}ska (\cite{Stas2002}) 
presented photoionization
models for a Milky Way Bulge PN with a central star temperature of 39\,000K
that can be equally well fit by a lower abundance ([O/H]=-0.31) and by a 
higher abundance model ([O/H]=+0.39)\footnote{Attempts to reproduce the
photoionization fits of Stasi\'{n}ska (\cite{Stas2002}) using Cloudy 08.01 were 
not entirely successful. The lower metallicity 
model could be satisfactorily reproduced with a slightly higher $T_{eff}$ of
43 kK and lower O of [O/H]=-1.02; however the higher metallicity 
model required [O/H]=+0.86 and at such high abundance the Cloudy
models were very sensitive to extremely small changes in abundances of O, Ne, N
and of the nebular size and density. From this numerical experiment, it is 
tentatively concluded that such extreme conditions are rare.}. 95\% of the PNe 
observed in NGC~5128 appear to have $T_{eff} \geq$ 60kK, so the likelihood 
of a double-valued abundance is very low. However since the lowest 
$T_{eff}$ values derived in the Cloudy modelling (Tab. 7) are $\sim$50kK, 
then uniqueness may be a more significant concern for these few objects.

A concerted attempt was made to fit the spectra of the PNe with the lowest 
derived T$_{eff}$ by low and high metallicity (Z) models. Here there is the 
liberty to change the stellar temperature since the He~II 4686/H$\beta$ 
ratio can be taken as an upper limit based on the errors on local lines. 
Two PN were found that could not be distinguished in terms of lower and 
higher Z Cloudy models. In the case of F56\#12b (a class b spectrum with the
[O~II] 
line detected but not He~II), the higher Z solution with 12+log(O/H) of 
8.55 for a BB $T_{eff}$ of 49 kK could not be distinguished from a model 
with $T_{eff}$ of 73 kK and 12+log(O/H) of 8.06 if the density 
was decreased from $4.0\times10^{4}$ to $1.8\times10^{4}$ cm$^{-3}$. 
For F42\#3 (class c spectrum without [O~II] and He~II lines detected), a higher Z
solution with 12+log(O/H) of 8.44 for $T_{eff}$ of 50 kK was indistinguishable from 
that with 12+log(O/H) of 8.01 but $T_{eff}$ of 73kK, for the same density in the shell. 
For the class c PN, an attempt was made to match each spectrum by a forced lower 
Z model but no convincing matches could be found; nevertheless, one may exist given the rather weak
constraints implied by these lower quality spectra. Conservatively, the higher 
abundance solutions were adopted (in Tab. \ref{Cloumods}) on the grounds that
they do not  differ strongly from the values for other PNe at similar radii.

The two PNe with alternate low O/H values are not improbable and are comparable to
the abundances of PNe in dwarf galaxies (e.g. Dopita \& Meatheringham, 
\cite{Dopita91} and Leisy \& Dennefeld, \cite{Leisy06}); F42\#4 also shows a 
similarly low value of O/H but has a well-detected spectrum. In the context of a 
giant elliptical galaxy, such low abundance values are surprizing because they
imply low metallicity progenitor stars,. Radial velocities of these PNe 
(available for F56\#12b and F42\#4 from Peng et al. \cite{Peng04}) do not show 
'peculiar' velocities, being within 1$\sigma$ of the mean velocity of NGC~5128. 
There is no indication of these PNe being obvious 'halo' objects or of resulting 
from a recent low-mass galaxy interloper not yet in dynamical equilibrium with the 
parent galaxy.

The abundances of the individual PN from Tab. \ref{Cloumods} were combined 
and compared to the abundances derived from the integrated region spectra with 
empirically determined temperatures and ICFs presented in Tab. \ref{Regabunds}.
The means were determined by weighting the individual abundances by 
m$_{5007A}$. For the combined region F56, the weighted mean abundances 
from Tab.  \ref{Cloumods} for O agreed to within 0.01 dex and for F34 to 
within 0.06 dex (weighted means of the log and the fractional O/H abundances 
were calculated with similar results). For field F42, the discrepancy 
on the weighted mean of the log O abundance was larger, with a value up to 
0.22 dex lower than the value in Tab. \ref{Regabunds}. The difference for 
F42 is quite striking in this comparison; the electron temperature was 
measured to be $\sim$1\,000K lower than for the other two regions. It is
suggested that the value of T$_e$ was underestimated in this
merged spectrum due to a poorly determined [O~III]4363\AA\ flux.

A further check on the integrated region spectra was performed by
running photoionization models for the region spectra presented 
in Tab. \ref{RegDerFlux}. The abundances were compared to the empirical 
abundances derived using ICFs (Tab. \ref{Regabunds}). Satisfactory 
Cloudy models could not be fitted without departing from the simplifying 
assumptions used for the individual PN models. In particular, when the 
stellar temperature was chosen from the He~II/H$\beta$ ratio, this could 
not be made compatible with the electron temperature derived from the 
[O~III]5007/4363\AA\ ratio. If He~II4686\AA\ matched the observations 
within the errors, then [O~III]4363\AA\ was predicted much too
strong (higher T$_e$); when [O~III]4363\AA\ matched the model, 
He~II was modelled too low (lower $T_{eff}$). The addition of modest 
amounts of dust inside the ionized region of the photoionization model to 
provide extra heating through photoelectric emission from the 
grains did not resolve this discrepancy. However allowing for 
these modelling issues, the range of O/H was close (within 0.1 dex) 
to the empirical values listed in Tab. \ref{Regabunds} for the sum of PNe 
in fields F56 and F34, but only within 0.2 dex for F42. 
These demonstrations show that summing PN spectra provides a useful 
indicator of the mean O abundance in cases when the line detections 
have low S/N, such as for PNe in more distant galaxies. This
conclusion is subject to the condition that the PN abundances do not 
differ strongly, as in this case. M\'{e}ndez 
et al. (\cite{Mendez05}) have presented lower limits on O and Ne abundances 
based on just such a summed spectrum for 14 PNe in NGC~4697 (an elliptical 
at 11 Mpc). 

\subsection{Towards a comparative study of models}	

A number of Cloudy photoionization models were run in order to arrive 
at an adopted model for each PN matched in Tab. \ref{Cloumods}. The 
models presented cannot be claimed to be unique but represent 
a feasible match to the spectra in terms of the abundances and 
physical conditions within the restricted set of parameters (black body
atmosphere for the stars, linear density law, etc). Magrini et al. 
(\cite{Magrini}) and JC99 discussed the accuracy of photoionization
models to their PN spectra. In general the ``a" quality spectra of the NGC~5128
PN approach those of the fainter sample of the much closer PNe observed in M33 
and M31 by these authors. Simply adopting the range of models within
the statistical errors on the line ratios underestimates the real uncertainties
on the abundance determinations. For a selection of models, independent Cloudy
runs were performed by two of the authors and the results compared (c.f.
Jacoby et al. \cite{Jacoby97}) and with the ICF determinations in the cases
where [O~III]4363\AA\ was detected (using an adopted N$_e$ value of 
5000 cm$^{-3}$). While a full error analysis is outside the scope of this
work, a limited investigation was attempted. A range of parameter
values around the adopted ones were explored to determine the
sensitivity of the models, given the errors on the spectra. 

It would be very time consuming to perform an investigation of
the likely parameter range for all PN in Tab. \ref{Cloumods}, so two 
PN were selected. F56\#2 was chosen as a high S/N case with both 
[O~III] 4363\AA\ and He~II 4686\AA\ well detected (a class ``a" spectrum 
in Table \ref{Cloumods}) and a high stellar temperature. F34\#1 was 
chosen as a PN with lower S/N and only a marginal detection of He~II 
4686\AA\ (class ``b" in Table \ref{Cloumods}) and a lower stellar 
temperature and representing the typical average quality of spectra. 
Cloudy models were run for these two exemplars varying a 
number of parameters about the adopted solution in Tab. 
\ref{Cloumods}: O abundance varied by $\pm$0.2 dex; stellar black 
body temperature by 10-15\% (20\,000K for F56\#2, with $T_{BB}$ 
155\,000K and 10\,000 K for F34\#1, with $T_{BB}$ 71\,000 K); the 
density by a factor of two; and a model atmosphere rather than a 
black body. For each modified parameter set, Cloudy models were run 
to match the spectra within errors as far as possible by freely 
varying all the other, non-displaced, parameters. Even this 
limited analysis involved running more than 180 separate Cloudy 
models.

The results of the comparison of Cloudy models are summarised in Tables 
\ref{VarF56} and \ref{VarF34}. The third columns list the dereddened 
spectrum from Tab. \ref{DeredFlux} and the 4th column the spectrum 
resulting from the adopted model given in Tab. \ref{Cloumods}. The 
successive columns list the resulting spectra from the incremented 
parameters. The lower half of each table lists the stellar temperature, 
density and abundances corresponding to the spectrum in the upper part 
of the same column; values in column 4 being identical to those in 
Tab. \ref{Cloumods}. An overall merit factor, $FoM$, of the match of 
the model spectrum with the observed spectrum, taking account of the 
S/N of the fluxes, is defined by:
\begin{equation}
FoM = \sum_{i=1}^{n} |((F_{obs}-F_{mod})/F_{obs})|\times (F_{obs}/Err_{obs}))
\end{equation}
where $F_{obs}$ and $Err_{obs}$ are respectively the dereddened flux 
and error and $F_{mod}$ the model flux; the summation is taken over the 
number of emission lines. Since some lines are in fixed ratio, such as 
H lines, a subset of the lines was adopted to dispense with some 
redundant information (except for the more important lines such as 
H$\alpha$ and H$\beta$, the weak He~I lines and the [O III] 4959\AA\ 
line). 

Table \ref{VarF56} presents the comparative models for the bright 
PN F56\#2 (5601 in Hui et al. \cite{Hui93a}). The comparison of the observed 
and model spectra makes it clear that the He~I 5876\AA\ line is strongly 
under-estimated, most probably because of the poor removal of the
strong Na~I telluric emission. In general satisfactory models could be
found, except for the cases of O - 0.2 abundance (column 6 of
Table \ref{VarF56}) and $T_{BB}$ - 20\,000 K (column 8).
What constituted a satisfactory match to the spectrum was not simply 
given by the FoM value, but certain line ratios, which act as
important diagnostics of nebular conditions, e.g. He~II/H$\beta$,
[O~III] 5007\AA/H$\beta$ and [O~III] 5007/4363\AA), were given higher
weight in assessing the quality of the match. It is clear that 
a range of O $\pm$0.2 produces model spectra in which the important
[O III] lines are not well-matched, and thus representative 
errors on O in the range 0.1 to 0.15 are suggested. Similarly
the range of stellar temperature of $\pm$20\,000 K and density
varying by a factor 2 also over-estimate the allowed range of some of 
the line ratios. However it is clear from these parameter ranges 
that the error bars are not necessarily symmetric. H, He and PG1159 
atmospheres from Rauch et al. \cite{Rauch03} were input to Cloudy and 
although the choice of temperatures is not very extensive, a satisfactory 
fit with the PG1159 atmosphere for 170\,000 K was found. However this 
should not be considered as a best estimate since a comprehensive range 
of temperature and gravity was not tested.

The comparative models for the fainter PN F34\#1 (2906 in Hui et al. 
\cite{Hui93a}, $m_{5007A}$=24.96) are given in Tab. \ref{VarF34}. Here
the number of lines is less than F56\#2, since [O~III]4363\AA\, and no
He~I lines were detected and He~II is a limit. The [O~II] 7325\AA\
line is clearly under-estimated in the measured spectrum by comparison with
the model spectra. Satisfactory matches could generally be obtained with 
the ranges of parameters listed. The models with a range of O of 
$\pm$0.2 is clearly accommodated by the data. Only the lowering of the 
stellar temperature by 10\,000 K (column 8 of \ref{VarF34}) produced an
unsatisfactory fit. A PG1159 model atmosphere (Rauch \cite{Rauch03}) 
of 70\,000K produced a good fit (column 11), but the 70\,000 K He 
model produced a similar fit with an O abundance lower by 0.40 dex 
than the black body. 

From this comparison of observed and model spectra with a higher and a 
lower S/N, we conclude that the typical dependence of errors on the
choice of Cloudy models are generally in the range of 0.15 - 0.20 for 
the O abundance, around 10-15\% for He and around 0.2 for Ne and N.
These values are similar to those found by Magrini et al.
\cite{Magrini} comparing abundances from Cloudy models with those of
the ICF method. These representative abundance errors are shown
in plots, such as Figs. \ref{abund_rels} and \ref{oh_grad}.

\begin{table*}
\caption[]{Comparative Cloudy photoionization modelling of F56\#2}
\label{VarF56}

\centering 
\begin{tabular}{ l | c | r | r | r | r | r | r | r | r | r }
\hline\hline

 Species & $\lambda$ (\AA)     & Flux \&  & Model & Model & Model & Model & Model & Model & Model & Model~~~~ \\   
         &                     & error    &       & O + 0.2 & O - 0.2 & T$_{BB}$ + 20 kK & T$_{BB}$ - 20 kK 
& N$_e$ $\times$ 2.0 & N$_e$ $\times$ 0.5 & Mod.atmos. \\
\hline
~[O~II]                 & 3727 &   56 $\pm$ 13 &    59 &    58 &    47 &    53 &    59 &    32 &    53 &    57 \\
~[Ne III]               & 3868 &  124 $\pm$ 13 &   124 &   127 &   128 &   122 &   123 &   126 &   125 &   125 \\
~[O~III]                & 4363 &   28 $\pm$ ~8 &    24 &    15 &    27 &    26 &    22 &    22 &    22 &    27 \\
~He~I                   & 4471 &    6 $\pm$ ~3 &     6 &     6 &     6 &     3 &    11 &     6 &     9 &     6 \\
~He~II                  & 4686 &   39 $\pm$ ~8 &    39 &    38 &    34 &    46 &    30 &    37 &    30 &    38 \\
~H$\beta$               & 4861 &  100 $\pm$ ~0 &   100 &   100 &   100 &   100 &   100 &   100 &   100 &   100 \\
~[O~III]                & 4959 &  457 $\pm$ 31 &   449 &   448 &   398 &   451 &   471 &   454 &   458 &   456 \\
~[O~III]                & 5007 & 1367 $\pm$ 90 &  1350 &  1348 &  1199 &  1358 &  1417 &  1367 &  1377 &  1371 \\
~He~I                   & 5876 &    6 $\pm$ ~3 &    18 &    18 &    19 &    11 &    33 &    19 &    28 &    20 \\
~H$\alpha$              & 6562 &  281 $\pm$ 19 &   286 &   285 &   290 &   286 &   286 &   284 &   285 &   288 \\
~[N~II]                 & 6583 &   55 $\pm$ ~5 &    54 &    57 &    58 &    53 &    53 &    54 &    53 &    57 \\
~He~I                   & 6678 &    7 $\pm$ ~4 &     4 &     5 &     4 &     2 &     8 &     4 &     7 &     5 \\
~[Ar~III]               & 7133 &    8 $\pm$ ~3 &     9 &     6 &    10 &     8 &     9 &     7 &     9 &     8 \\
~[O~II]                 & 7325 &   14 $\pm$ ~4 &    16 &    14 &    13 &    19 &    14 &    15 &     9 &    16 \\
                        &      &               &       &       &       &       &       &       &       &      \\
\hline
% Parameter             &      &               &   Fit & O+0.2 & O-0.2 & T+20K & T-20K &  Ne*2 &  Ne/2 &   Mod. \\
Model input             &      &               &        &        &        &        &        &        &        &         \\
\hline
~T$_{BB}$(kK)           &      &               &   155 & 152.5 &   150 &   175 &   135 &  150  & 132.5 &    170$^{1}$ \\
~N$_H$ (cm$^{-3}$)      &      &               & 10000 & 10000 & 10000 & 12000 &  8000 & 20000 &  5000 &  10000 \\
~12+log(He)             &      &               & 11.15 & 11.15 & 11.20 & 11.00 & 11.35 & 11.15 & 11.30 &  11.18 \\
~12+log(C)              &      &               &  8.40 &  9.20 &  6.60 &  8.40 &  8.40 &  8.70 &  8.40 &   8.20 \\ 
~12+log(N)              &      &               &  7.85 &  8.02 &  7.84 &  7.85 &  7.88 &  8.01 &  8.05 &   7.85 \\
~12+log(O)              &      &               &  8.43 &  8.63 &  8.23 &  8.48 &  8.43 &  8.49 &  8.40 &   8.39 \\
~12+log(Ne)             &      &               &  7.72 &  7.96 &  7.57 &  7.75 &  7.70 &  7.78 &  7.70 &   7.67 \\ 
~12+log(Ar)             &      &               &  6.00 &  6.00 &  6.15 &  6.00 &  6.00 &  6.00 &  6.00 &   5.95 \\ 
                        &      &               &       &       &       &       &       &       &       &      \\
\hline
~FoM                    &      &               &   7.53 &   8.30 &  13.46 &   7.34 &  14.79 &   9.06 &  12.90 &    8.56 \\
% Model No.               &      &               &   604 &   310 &  1135 &   153 &   143 &   192 &   175 &     58 \\
\hline
\end{tabular}
\tablefoot{
$^{1}$ Model atmosphere: PG1159 atmosphere with T$_{*}$ = 170\,000 K and  $log~g$ = 7.00 from Rauch \cite{Rauch03} \\ 
S abundance fixed at 12+log(S/H) = 6.55 \\
}
\end{table*}

\begin{table*}
\caption[]{Comparative Cloudy photoionization modelling of F34\#1}
\label{VarF34}

\centering 
\begin{tabular}{ l | c | r | r | r | r | r | r | r | r | r }
\hline\hline

 Species & $\lambda$ (\AA)     & Flux \&  & Model & Model & Model & Model & Model & Model & Model & Model~~~~ \\   
         &                     & error    &       & O + 0.2 & O - 0.2 & T$_{BB}$ + 10 kK & T$_{BB}$ - 10 kK 
& N$_e$ $\times$ 2.0 & N$_e$ $\times$ 0.5 & Mod.atmos. \\
\hline
~[O~II]                 & 3727 &   86 $\pm$ ~9 &    89 &    85 &    88 &    87 &    60 &    49 &   144 &    85 \\
~[Ne III]               & 3868 &   93 $\pm$ ~6 &    91 &    92 &    93 &    92 &    95 &    92 &    94 &    96 \\
~He~II                  & 4686 &    2 $\pm$ ~2 &     2 &     2 &     4 &     4 &     1 &     2 &     4 &     0 \\
~H$\beta$               & 4861 &  100 $\pm$ ~0 &   100 &   100 &   100 &   100 &   100 &   100 &   100 &   100 \\
~[O~III]                & 4959 &  393 $\pm$ 15 &   384 &   391 &   384 &   393 &   248 &   385 &   387 &   390 \\
~[O~III]                & 5007 & 1164 $\pm$ 41 &  1156 &  1177 &  1157 &  1183 &   747 &  1157 &  1166 &  1174 \\
~H$\alpha$              & 6562 &  283 $\pm$ 12 &   288 &   290 &   287 &   286 &   286 &   287 &   288 &   287 \\
~[N~II]                 & 6583 &  104 $\pm$ ~6 &   104 &   104 &   104 &   104 &   103 &   104 &   105 &   103 \\
~[Ar~III]               & 7133 &    4 $\pm$ ~3 &     3 &     4 &     4 &     4 &     3 &     3 &     4 &     4 \\
~[O~II]                 & 7325 &    4 $\pm$ ~3 &    17 &    16 &    17 &    17 &     8 &    16 &    17 &    14 \\
                        &      &               &       &       &       &       &       &       &       &      \\
\hline
% Parameter             &      &               &   Fit & O+0.2 & O-0.2 & T+10K & T-10K &  Ne*2 &  Ne/2 &   Mod. \\
Model input             &      &               &        &        &        &        &        &        &        &         \\
\hline
~T$_{BB}$(kK)           &      &               &    71 &    68 &    78 &    81 &    61 &  67.5 &    80 &     70$^{1}$ \\
~N$_H$ (cm$^{-3}$)      &      &               &  8500 & 10000 &  7500 &  8500 &  8500 & 17000 &  4250 &   7000 \\
~12+log(He/H)           &      &               & 11.06 & 11.06 & 11.06 & 10.90 & 11.30 & 11.06 & 10.90 &  11.20 \\
~12+log(C)              &      &               &  8.60 &  8.50 &  8.40 &  8.80 &  8.50 &  8.60 &  8.30 &   8.20 \\ 
~12+log(N)              &      &               &  8.25 &  8.38 &  8.13 &  8.16 &  8.50 &  8.40 &  8.03 &   8.18 \\
~12+log(O)              &      &               &  8.73 &  8.93 &  8.53 &  8.60 &  9.00 &  8.78 &  8.54 &   8.60 \\
~12+log(Ne)             &      &               &  8.00 &  8.23 &  7.78 &  7.84 &  8.60 &  8.07 &  7.77 &   7.88 \\ 
~12+log(Ar)             &      &               &  5.65 &  5.85 &  5.55 &  5.60 &  5.90 &  6.65 &  5.55 &   5.60 \\ 
                        &      &               &       &       &       &       &       &       &       &      \\
\hline
~FoM                    &      &               &   6.44 &   5.79 &   6.48 &   6.74 &  26.62 &  10.20 &  12.90 &     5.30 \\
% Model No.               &      &               &    10 &    26 &    36 &   47 &   526 &   1037 &   853 &    139 \\
\hline
\end{tabular}
\tablefoot{
$^{1}$ Model atmosphere: PG1159 atmosphere with T$_{*}$ = 70\,000 K and  $log~g$ = 7.00 from Rauch \cite{Rauch03} \\ 
S abundance fixed at 12+log(S/H) = 6.55 \\
}
\end{table*}

\section{Discussion}

\subsection{Individual PN spectra and abundances}
A sample of 51 PNe in NGC~5128 have been observed over a range of 
projected galactocentric distances from 1.7 to 18.9 kpc and 
covering the PN luminosity function from the brightest known PN in 
NGC~5128 (PN 5601, $m_{5007A} = 23.51$) to 4.1 mags fainter. The 
mean 5007\AA/H$\beta$ ratio is 11.00 $\pm$ 0.55 (r.m.s on the 
mean) for the 42 spectra with the highest S/N. The extrema of 
the 5007\AA/H$\beta$ values for this subset are 6.0 
($\pm$ 0.67) to 18.8 ($\pm$3.7); the value of this 
ratio is very sensitive to the subtraction of the 
underlying continuum and the variation of the H$\beta$ 
absorption line along the slitlet. The mean logarithmic 
extinction correction at 
H$\beta$ is 0.56 $\pm$ 0.20. The overall Galactic
extinction to NGC~5128 is E$_{B-V}$=0.123 (Burstein \& Heiles
\cite{BurHei}) or E$_{B-V}$=0.115 (Schlegel et al 
\cite{Schlegel}), the latter is equivalent to c=0.17 for a 
Galactic extinction law with R=3.2 (Seaton \cite{Seaton}).
The lowest values of extinction measured in the PNe
(Tab. \ref{DeredFlux}) are compatible with no local 
extinction in NGC~5128. The histogram of c values peaks at 0.45 
($E_{B-V}$ = 0.30) and shows no obvious trend with radial offset 
beyond 200$''$ from the nucleus; the five highest values occur 
within a radius of 200$''$ (3.7 kpc). The frequency of higher c 
values at lower radii is probably due to the high line of sight 
extinction in the vicinity of the dust lane, rather than any 
intrinsic PN property (e.g. high intrinsic dust content associated with 
young PNe, such as NGC~7027;
Zhang et al. \cite{Zhang}). 

The ratio of 5007\AA/H$\beta$ does not show any significant
overall gradient with projected radial offset from the galaxy centre;
the highest values lie in field F42 where the stellar continuum
is strongest and removal of the underlying H$\beta$ is the most
critical. There is no evidence for a gradient in [Ne~III]/[O~III]
implying a constant O$^{++}$/Ne$^{++}$ ratio. If there is 
any O enrichment, e.g. during the third dredge-up (P\'{e}quignot et al. 
\cite{Pequ}), or depletion (through the ON cycle, Leisy \& Dennefeld
\cite{Leisy06}) occurring in the nebular
envelopes of the PNe, then this constant value of the 
ionic ratio implies that any alteration in O abundance
must be accompanied by a corresponding change in Ne. Since O-Ne 
correlation is not predicted by AGB evolution (Karakas \& Lattanzio 
\cite{KarLat}), the rather constant O/Ne ratio implies that O 
enrichment/depletion is not an important effect. 

The mean 12 + log(O/H) is 8.52 $\pm$ 0.03 (median value 8.48) and the 
range of [O/H] is $-$0.66 to $+$0.31. The mean O/H abundances inside
and outside 10 kpc are identical. Of the three PN with the 
lowest O/H, two occur close to the nucleus (F42\#4 and F42\#10) 
and the other is at 19.5 kpc (F34\#2) and all have low [O~III]/H$\beta$,
without notably strong [O~II]; F34\#2 has lower N/H. The PN 
with the highest O/H occur in fields 42 and 
56 and are distinguished by very high [O~III]5007/H$\beta$
but with large errors. The problems in removing the underlying
stellar continuum from slit measurements can contribute to 
large uncertainty on the line fluxes for fainter nebulae
in the central regions (see the discussion in Roth 
(\cite{Roth06})), and very high [O~III]/H$\beta$
of $>$ 15 must be viewed with caution since the H$\beta$ may
be considerably underestimated. 

The mean N/O ratio is 0.51 $\pm$ 0.06. Four PNe were found with high N/O 
ratio (F56\#8, F56\#12b, F56\#18 and F42\#4) but only F56\#12b and F56\#18
have He/H and N/O which could be classified as characteristic
of Type I nebulae (Peimbert \cite{Peimbert}). Since F56\#12b
was modelled by a lower temperature central star (the lowest of
all the PN modelled, Tab. \ref{Cloumods}), it may be that 
optical depth effects and the assumption of a black body may
yield misleading results. F56\#18 is perhaps a better candidate for
a Type I nebula, having a relatively high stellar temperature
and a higher reddening, but this PN does not possess the highest
luminosity central star. The brightest PN, and incidentally
the nebula with the highest luminosity star (Tab. \ref{Cloumods}),
has N/O of 0.12 and shows no elevation of this ratio, compared
for example to the mean Milky Way value of 0.47 
(Kingsburgh \& Barlow \cite{KiBa}).
The shorter timescales for PN evolution of Type I PN make them minor 
contributors to a PN population luminosity function, but 
they contribute about $\sim$10\% of Milky Way PN by number. 

\begin{figure}
\centering
\resizebox{\hsize}{!}{\includegraphics[width=10.0cm,angle=-90]{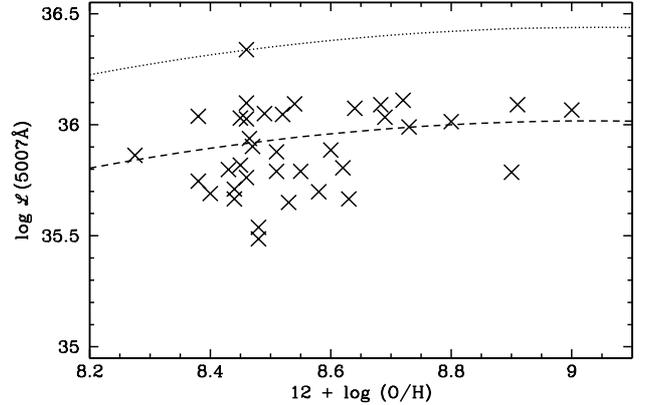}}
\caption{The dependence of the oxygen abundance of the 
NGC~5128 PN sample from the Cloudy photoionization modelling
(Tab. \ref{Cloumods}) on the observed
[O~III] 5007\AA\ luminosity (uncorrected for
reddening) from the 5007\AA\ photometry of Hui et al. \cite{Hui93b}. 
The dotted line shows the theoretical
relation of Dopita et al. (\cite{DopJac}) scaled to the
brightest PN, while the dashed line shows a second order
polynomial fit to all the data points.
}
\label{o3lum}
\end{figure}

The mean oxygen abundances in NGC~5128 PNe are intermediate 
between values for the LMC (12 + log(O/H) mean 8.4 (e.g. 
Leisy \&  Dennefeld\cite{Leisy06}), and that of the Milky Way
(mean 8.68, Kingsburgh \& Barlow \cite{KiBa}). A plot of the
dependence of the [O~III]5007\AA\ luminosity on 12+log(O/H)
(Fig. \ref{o3lum}) shows a tendency to increase with the 
oxygen abundance, in line with the theoretical relation of 
of Dopita et al. (\cite{DopJac}) as observationally transformed
by Richer (\cite{Rich93}; see Figure 2 and equation 1). A second 
order fit to the observed points demonstrates a comparable 
dependence of [O~III] luminosity on (O/H) to the theoretical 
relation and is shown in Fig. \ref{o3lum} by a dashed line.

\begin{figure}
\begin{center}
\resizebox{\hsize}{!}
{\includegraphics[width=10.0cm,angle=-90]{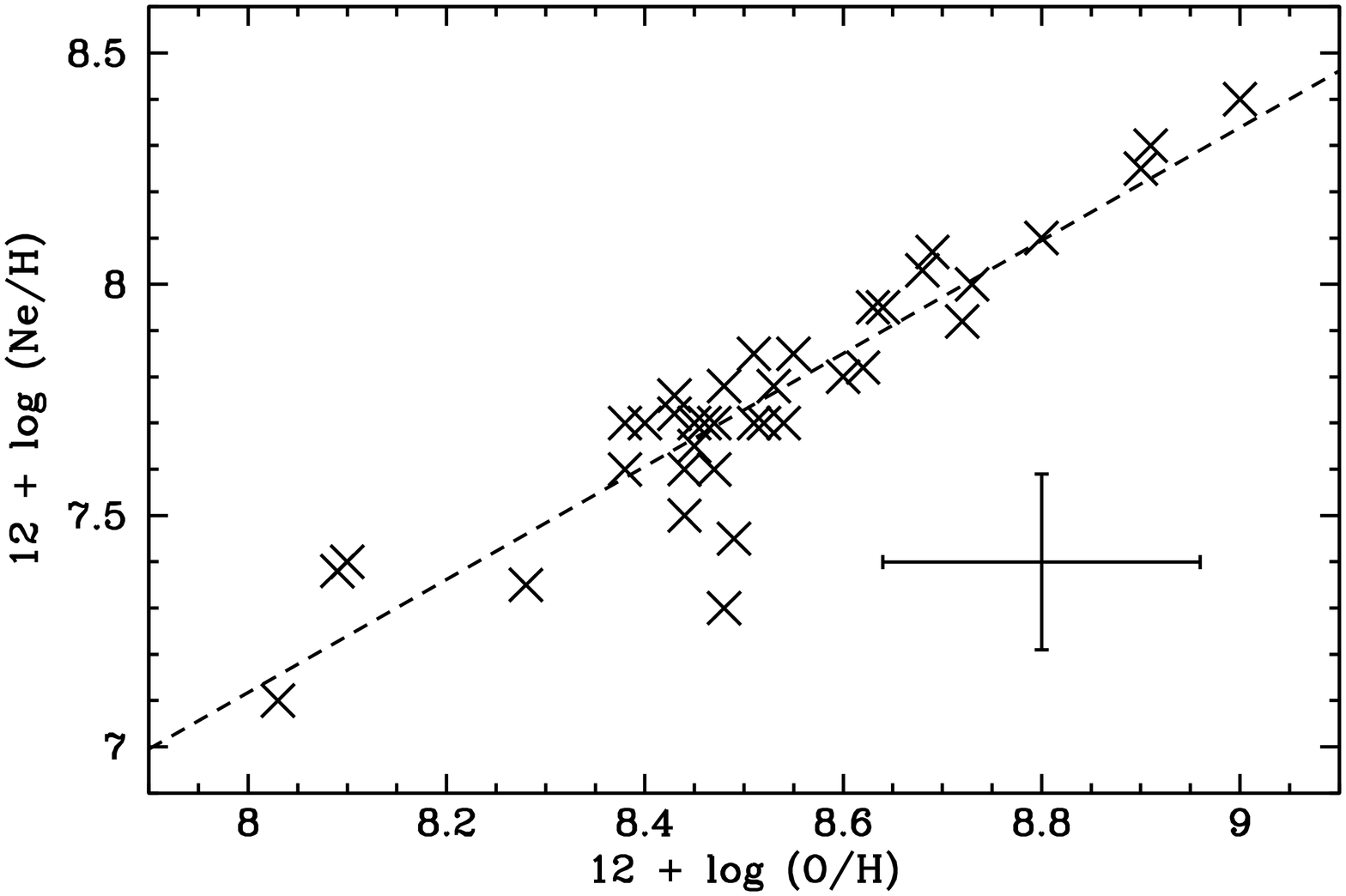}}
\hspace{1.0truecm}
\resizebox{\hsize}{!}{\includegraphics[width=10.0cm,angle=-90]{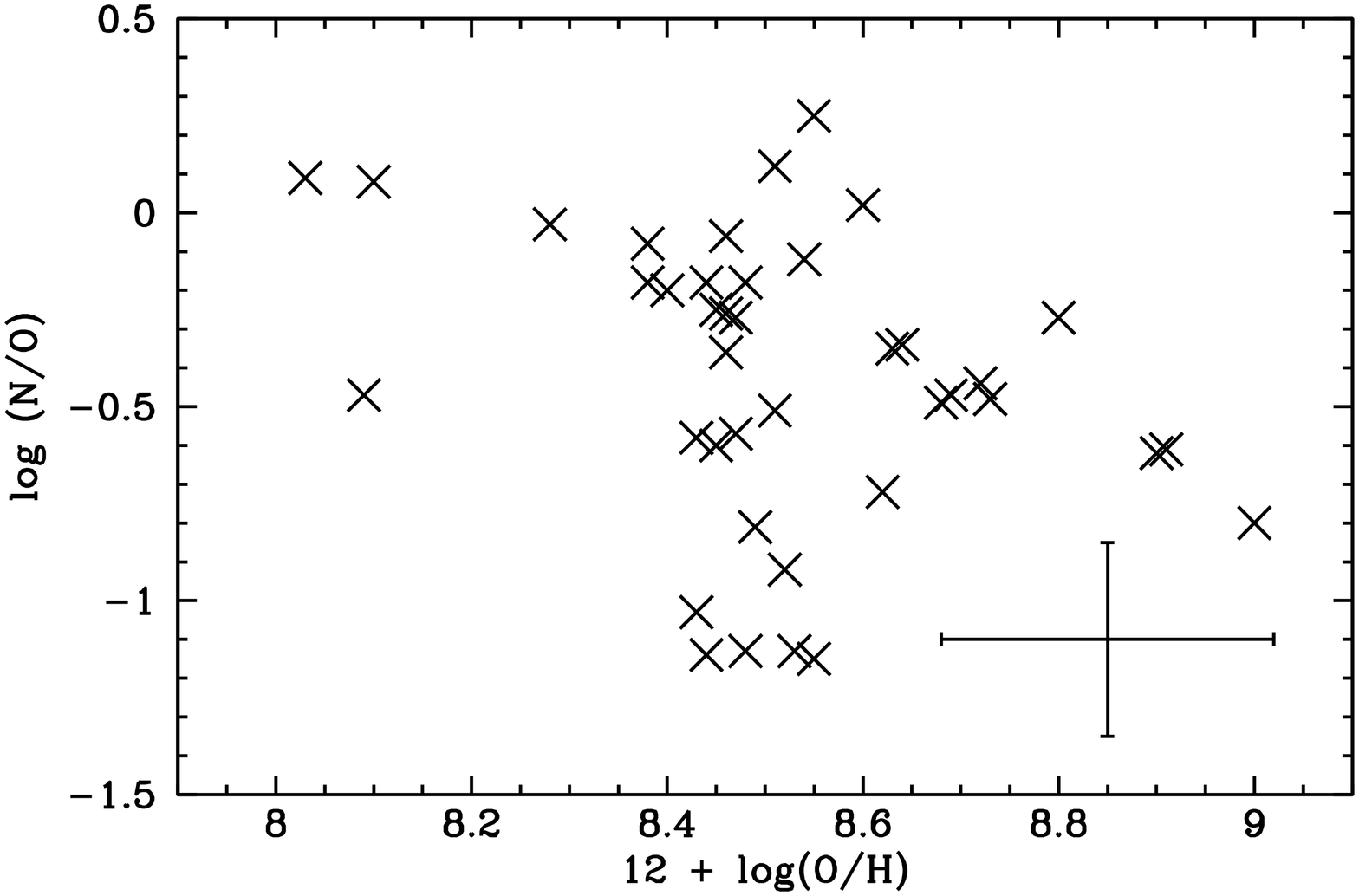}
}
\end{center}
\caption{The relationship between log (Ne/H) and log (O/H) is shown (upper) and 
for Log(N/O) against log (O/H) (lower) for the individual NGC~5128 PN 
abundances tabulated in Tab. \ref{Cloumods}. The representative error 
bars are indicated.}
\label{abund_rels}
\end{figure}

The region summed spectra provide good detections of a single
ionization species of S and Ar (Tab. \ref{RegDerFlux}) and lines 
of these species were detected in a number of the brightest PNe 
(Tab. \ref{ObsFlux} and \ref{DeredFlux}) allowing useful 
indications of S/H and Ar/H (Tab. \ref{Regabunds} and 
Tab. \ref{Cloumods}). S and Ar abundances serve an important 
purpose in that they are not considered to be affected by the 
AGB nucleosynthesis (e.g. Herwig \cite{Herwig}).
Whilst O is the element which is most easily measured in PN 
since it is the dominant coolant, it can be destroyed by CNO cycling
or synthesized during helium burning which impacts its usefulness as 
a metallicity indicator. P\'{e}quignot et al. (\cite{Pequ})
have found evidence for third dredge-up O enhancement in a PN in the
Sagittarius dwarf galaxy (12+log(O/H) = 8.36), but only at the level of 
$<$0.03 dex. By comparing the lock-step dependence of Ne and O in
PN samples, Richer \& McCall (\cite{Rich08}) conclude, in agreement 
with Karakas \& Lattanzio (\cite{KarLat}), that the majority of PNe do 
not show any changes in the O or Ne abundances across the AGB and PN 
transition. The Ne/H abundances are well correlated with
the O abundances for the NGC~5128 PNe as shown by
Fig. \ref{abund_rels} (slope 1.22$\pm$0.09, or 1.17$\pm$0.04 
excluding the 4 points furthest from the linear relation) This result is 
similar to the value found by Leisy \& Dennefeld 
(\cite{Leisy06}) for the LMC (slope 1.13).
The relation between N/O and O/H is also shown in Fig. \ref{abund_rels}.
There is considerable scatter with a weak but insignificant trend for 
an anti-correlation, as also found by Magrini et al. \cite{Magrini} 
for PNe in M33.
The mean Ar/H abundance (6.3 for 21 PNe) is indistinguishable from the 
mean for 70 PN in the Milky Way (Kingsburgh \& Barlow \cite{KiBa}). 
The mean S abundance (6.8 for 9 PNe) also appears to be similar
to the mean value from the Kingsburgh \& Barlow 
(\cite{KiBa}) sample. One is led to the surprising conclusion
that most of the PN abundances in NGC~5128 are
not significantly different from the mean values for Milky Way PNe. 

\subsection{PN 5601}
PN 5601 (F56\#2) is the brightest known PN and the 
best-studied individual PN in NGC~5128, and
indeed of all PN known beyond the 
Local Group. Walsh et al. (\cite{Walsh99}) measured long slit
observations of this PN amongst a few others and the major
difference between the spectrum presented in Tab. \ref{ObsFlux}
is the H$\alpha$/H$\beta$ ratio and the higher He~II/H$\beta$
ratio. The earlier observations were taken with a fixed slit 
position angle over a considerable range of parallactic angle 
and suffer from wavelength dependent flux loss from the slit.
This is demonstrated by the very low value of extinction derived.
The value of extinction derived for this nebula is now seen
to be quite large, but only slightly higher than the mean value for 
all the observed PN (E$_{B-V}$=0.32), giving no evidence that it 
is intrinsically dustier than the fainter PNe. There is some 
expectation from examples of Milky Way PNe, that young, high mass, 
optically thick PN (c.f. NGC~7027, Zhang et al. \cite{Zhang}; 
NGC~6302 Wright et al. \cite{Wright}) have higher intrinsic dust 
extinction. 

\subsection{Abundance gradient from PNe}
\begin{figure}
\centering
\resizebox{\hsize}{!}{\includegraphics[width=10.0cm,angle=-90]{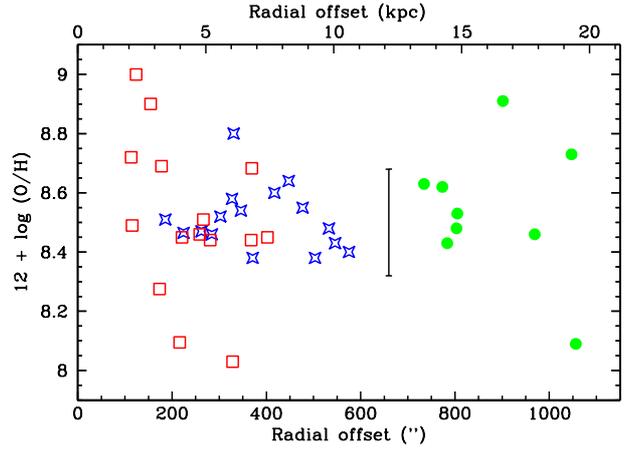}}
\caption{The radial dependence of the oxygen abundance
(12 + log(O/H)) from the observed PN modelled by Cloudy 
(Tab. \ref{Cloumods}). The points are coded for the three regions as in 
Fig. \ref{O3mags}: F56 by open stars (blue); F42 as open squares (red); 
F34 by filled circles (green). A representative error bar on the O 
abundance is shown.}
\label{oh_grad}
\end{figure}

One of the primary aims of this study was to investigate
any abundance gradient of the $\alpha$-elements, as revealed by
the PN spectra, and to compare it to gradients in the stellar
properties such as the metallicity distribution function.
A plot of the variation of O/H v. projected radial offset
(Fig. \ref{oh_grad}), for the PNe with O/H 
values from the Cloudy models (Tab. \ref{Cloumods}),
shows two particular features. First, the range of O/H values is
similar in the inner regions (projected radius, $\cal R$ $<$4 kpc) to the 
range in the outer regions ($>$13 kpc). Second, there is no 
evidence of a gradient in O/H. A least squares fit confirms this; 
the linear correlation
coefficient is  -0.07 for 40 points. 
Rather, there is a common mean value of 8.52 
for the inner and outer points (considered 
as less than and greater than $\cal R$ of 10 kpc). However, the highest
O/H values do occur at $\cal R$ $<$ 4 kpc, in the inner field 
F42.  Although our sample is not large, the result that there 
is no significant O/H gradient in NGC~5128 appears to be robust.

\begin{figure}
\centering
\resizebox{\hsize}{!}{\includegraphics[width=10.0cm,angle=0]{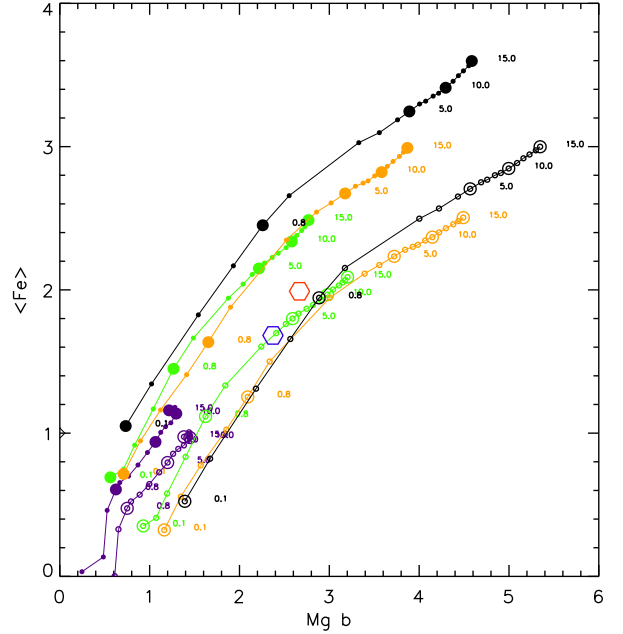}}
\caption{
The Lick index $<Fe>$ is plotted against Mg b for the 
integrated spectra of the two inner fields F42 (red lozenge) and F56 
(blue lozenge). Two families of tracks from Thomas et al. (2003) for [$\alpha$/Fe] 
of 0.3 (open symbols) and 0.0 (filled symbols) are shown for metallicity [Z] of -1.35, -0.33, 0.0 and 0.35 by 
lines with points. The lower set of four lines have  [$\alpha$/Fe] = 0.3, 
and the upper ones [$\alpha$/Fe]=0. Models with [Z]=0.35 are indicated 
in black, [Z]=0.0 in orange, [Z]=-0.33 in green and [Z]=-1.35 
in magenta. The ages in Gyr along each track are indicated.
} 
\label{O_Fe_rel}
\end{figure}

The distribution of [$\alpha$/Fe] v. [Fe] has an important role
in determining the enrichment history of a galaxy (e.g.
Matteucci \& Recchi \cite{MatRec}).
O, as representing the $\alpha$ elements, is primarily
a core collapse SN product released early in star formation
activity, whilst Fe, as the chief product of Type Ia SNe, 
is released over many Gyr. Figure 
\ref{O_Fe_rel} shows the Fe v. Mg b EW plot for the
integrated spectra of regions F42 and F56 with several
tracks from Thomas et al. (\cite{Thomas}) models for 
selected metallicity and [$\alpha$/Fe] ratio. To be compatible with 
Harris et al. (\cite{Harris00}), who calibrated their metallicities by assuming 
that their stars were as old as those in old Galactic globular
clusters, then the models must be compared at an age of 15 Gyr. 
In this case the datapoints for fields F42 and F56 lie close to the 
tracks of [$\alpha$/Fe] = 0.3, giving an [$\alpha$/Fe] ratio
of 0.25, and show metallicities [Z] around -0.6 (there are no tracks at 
metallicity intermediate between -0.33 and -1.35 in Thomas 
et al. \cite{Thomas}). If, however, the age of the stars were younger
(as indicated e.g. by Rejkuba et al. (\cite{Rejkuba05}, \cite{Rejkuba11})), 
the derived metallicities would be about 0.2 higher, giving an average 
value of -0.4. This would imply an oxygen abundance relative to solar of -0.15,
a value that is consistent with the mean O abundance of -0.17 with respect to 
the solar value (Scott et al. \cite{Scott}).

\begin{figure}
\centering
\resizebox{\hsize}{!}{\includegraphics[width=10.0cm,angle=-90]{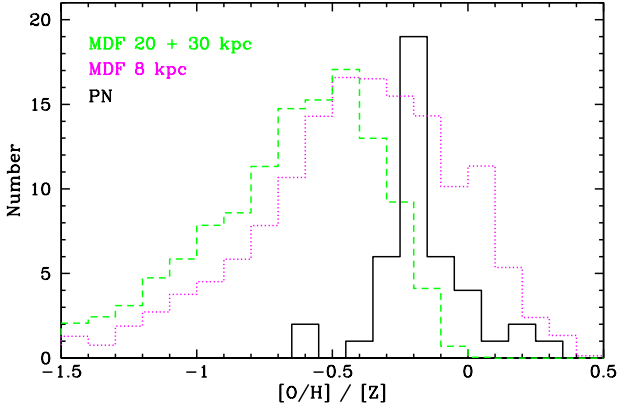}}
\caption{The histogram of the log O/H abundance relative to Solar
is plotted for all PN with O/H determined from Cloudy models (Tab.
\ref{Cloumods}) as a continuous line (black). For comparison the
stellar metallicity distribution function (MDF) from
Harris \& Harris (\cite{Harris02}) is plotted in two
regions: an inner region (8 kpc) dotted (magenta); a composite
of two outer regions at 20 and 30 kpc dashed (green).
}
\label{O_Z}
\end{figure}

These abundances can also be compared with the MDF 
from Harris \& Harris (\cite{Harris02}) in three fields 
at 7.6 kpc SW of the centre, at 20 kpc S and 29 kpc 
S (distances rescaled to 3.8 Mpc). Fig. \ref{O_Z} shows the
histogram of all the PN O/H abundances with respect to Solar
overplotted against the completeness corrected MDFs in the inner 
field (7.6 kpc) and two combined outer fields (Table 1 of Harris \& Harris
\cite{Harris02}). The PN sample covers the range around the inner and
outer fields of Harris et al. (\cite{Harris02}). A shift of -0.25 is 
required to align the peak of the PN O/H distribution with the stellar 
MDF in the inner field (magenta histogram). However it is strongly 
apparent that the distribution of (O/H) for the PNe is much narrower 
than for the stellar metallicity, so matching the distribution
peaks may not be a justifiable simplification. Splitting the PN into two samples 
with positions greater and less than 10 kpc shows a shift in the peak 
value of (O/H) of $\sim$-0.2 for the inner field and $\sim$-0.3 for the 
outer field, although the number of PNe with abundances is lower in the 
outer field (11). 
For the full PN sample, with an average O abundance of -0.17 compared 
to Solar, the total metallicity is then -0.4 assuming 
a mean [$\alpha$/Fe]=0.25. Harris et al. (\cite{Harris02}) find an 
average metallicity of -0.6 in their outer 
fields, which becomes -0.4 assuming an age of 8 Gyr; thus taken at 
face value, assuming the metallicity of the field stars is that
of the PN, the age of these stars in NGC~5128 is around 10 Gyr. 

Both methods -- matching observed spectral indices to 
stellar population models and comparison with photometrically
derived stellar metallicity distributions -- provide an average [O/Fe] 
ratio of 0.25 at a stellar metallicity (viz. [Fe/H]) of around 
-0.4, and an average age of 8 Gyr. This single value is obviously a 
simplification for the behaviour in a whole galaxy, but demonstrates 
that the [O/Fe] v. [Fe/H] value for NGC~5128 is within the range of 
$\alpha$-enhanced stellar metallicities for early type galaxies (c.f. 
Thomas et al. \cite{Thomas}). The modelling of the colour-magnitude 
diagram by Rejkuba et al. (\cite{Rejkuba11}) also independently favours
$\alpha$-enhanced stellar abundances in NGC~5128. 

However it can be argued that while all the stars are represented in the
integrated stellar spectra and photometric surveys, only a subset of these
stars become PNe. So, the mean values of the two samples of the PN [O/H] 
and the stellar [Fe/H] in Fig. \ref{O_Z} are not necessarily strictly 
comparable. In particular the PNe may derive from a younger population, 
on average. Only the higher Z stars produce observable PN, while the stellar 
continuum is comprised of stars of all metallicities. Overall the occurrence of PN is
highly delimited relative to all the stars in a galaxy and in addition the PN 
central star masses also have a very restricted range relative to all white 
dwarfs. Thus the PNe may only come from a young population; that would explain 
the tendency toward both higher Z (Fig. \ref{O_Z}) and higher central star 
masses (see Fig. \ref{HRdiag}). The old stars, which have lower Z on average, 
fail to produce PNe because their masses are too low.

Chiappini et al. (\cite{Chiap}) present a careful comparison of the 
spectroscopically determined abundances of Milky Way Bulge stars with 
Bulge planetary nebulae. Their sample of 166 PN in the Bulge with 
well-determined abundances shows a broad distribution of 12 + log(O/H) 
from about 8.2 to 8.9, with a mean of 8.57, rather similar to the mean 
value and range determined here in NGC~5128. However the O/H abundance collated 
from various spectroscopic studies of Bulge giant stars shows a lower 
value by $\sim$0.3 dex. Various suggestions are offered by Chiappini et al.
(\cite{Chiap}) for this discrepancy, but none appears to be conclusive. 
Comparing O from the Bulge PN with [O/Fe] from the stars, implies
[O$_{PN}$/Fe$_{Stars}$] $\sim$ -0.1. This value is not strictly 
comparable to that in NGC~5128 since [Fe/H]
is derived from photometry; spectroscopy of individual stars in NGC 5128
is ruled out until larger telescopes are available. Taken at face value 
the difference in [O/Fe] could imply that the formation history of 
NGC~5128 as a classical giant elliptical, is dissimilar to that of 
the Bulge; however the discrepancy in O abundances of Bulge PN and 
giants found by Chiappini et al. (\cite{Chiap}) weakens this
conclusion as does the large spread in stellar Fe abundances (in both
NGC~5128 and the Bulge).

\subsection{PN and the star formation history}
\begin{figure}
\centering
\resizebox{\hsize}{!}{\includegraphics[angle=-90]{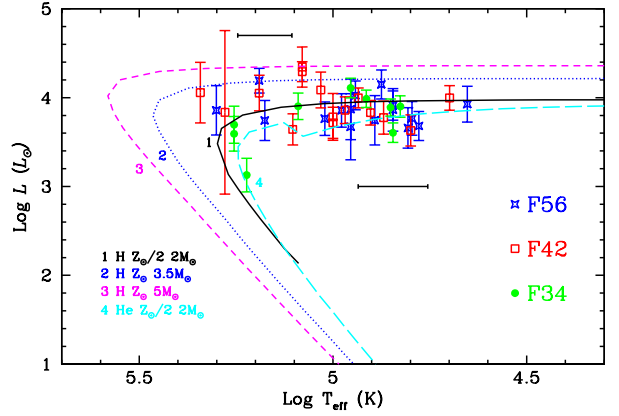}}
\caption{An HR diagram showing the black body log $L$ {\it v.} log $T_{eff}$ 
coordinates of the central stars of the PNe with spectra modelled 
by Cloudy (Tab. \ref{Cloumods}) and a selection of low mass H and He burning
model tracks from Vassiliadis \& Wood (\cite{VasWood}). The points are coded 
(and coloured) for the three regions (F42, F56 and F34) as in Fig. \ref{O3mags}. 
Representative error bars on temperature at higher and lower temperatures 
are shown; the error bars in $Log~L$ were computed from the errors on 
dereddened absolute H$\beta$ flux in Tab. \ref{DeredFlux}.
}
\label{HRdiag}
\end{figure}

One of the advantages of running Cloudy models to determine the
PN abundances is that the central star luminosities and temperatures
are derived. Since the PNe are optically thick, the match 
to the spectrum effectively provides the temperature and the $m_{5007\AA}$ 
photometry, combined with the reddening, provides the luminosity. 
Fig. \ref{HRdiag} shows the Log $L$, Log T$_{eff}$ points
for the 40 PNe modelled (Tab. \ref{Cloumods}). The log $L$ values
span a range of a factor 6, but are, not surprizingly, at the 
high end for PNe central stars, since only the top few magnitudes 
of the PNLF is explored by these data. No radial trend in the
central star luminosities is apparent, paralleling the lack of an
O abundance gradient (Fig. \ref{oh_grad}). A few evolutionary tracks 
from Vassiliadis \& Wood (\cite{VasWood}) for higher mass 
progenitor stars are shown on Fig. \ref{HRdiag}, 
although there is currently no information on whether the central 
stars are on the H or He burning track. For the H burning tracks the
bulk of the points match stars with masses above 2 M$_\odot$.

There are multiple routes in terms of stellar mass,
metallicity and mode of turn-off from the AGB (H or He burning
track) to reach the same PN luminosity, so relating 
a given PN in a galaxy to its progenitor star formation 
episode is non-trivial. Marigo et al. 
(\cite{Marigo}) have modelled effects of stellar populations
on the [O~III] cut-off of a population of PN, as measured for
the PNLF. They find that the peak [O~III] luminosity is emitted
by stars with initial (main sequence) masses of about 2.5 M$_\odot$
with a very strong dependence with age. PN from higher mass
progenitors, although being intrinsically luminous, evolve very 
rapidly and so make a minor contribution to the bright end of 
the PN population. In addition, Ciardullo \& Jacoby (\cite{Ciardullo99})
showed that the intrinsically massive and luminous progenitors are prolific
dust producers, and therefore self-extinct. Thus the most massive
progenitors will generally appear as rare and faint PNe in a galactic sample. 
Ciardullo et al. (\cite{Ciardullo05}) 
suggest that the progenitors of the most luminous PNe in galaxies, 
if they are not formed from stars younger than $\sim$1 Gyr,
may occur predominantly in binary systems.

\begin{figure}
\centering
\resizebox{\hsize}{!}{\includegraphics[width=10.0cm,angle=-90]{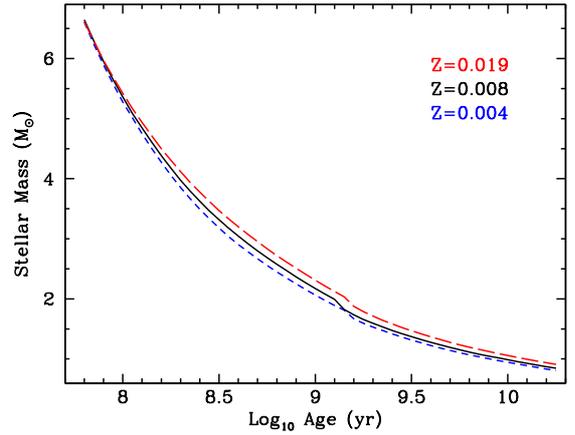}}
\caption{
The age at which a star terminates the AGB is shown as a 
function of the initial main sequence stellar mass for 
three metallicities of 0.019 (long-dashed line, red), 0.008 (Solar metallicity;
continous line, black) and 0.004 (approximately LMC metallicity; short-dashed line, blue). The curves are derived from the stellar evolution tracks of Girardi et al. 
(\cite{Girardi}).
}
\label{AGBage}
\end{figure}

In principle the calibration of PN luminosity v. stellar progenitor
mass can be used to identify the age of the starburst giving rise to
an observed PN population. If indeed
the bulk of the PN observed in this study arise in progenitors
of about 2.5 M$_\odot$, then from the grid of evolutionary tracks of
low mass stars of Girardi et al. (\cite{Girardi}) one can infer
a progenitor age of 0.7 Gyr (see Fig. \ref{AGBage} derived from
the Girardi et al. tracks for $[Z/Z_\odot]$ 0.0, -0.4 and -0.7 and 
the stellar age to AGB termination). Such an age would place the 
PN in a very recent star formation episode. This seems rather 
unlikely since there is little evidence for an extensive 
stellar population so young. A young (2-4 Gyr) minority (20-30\%) 
component was inferred from the HST stellar photometric studies of
Rejkuba et al. (\cite{Rejkuba05} and \cite{Rejkuba11}), but not
with stars younger than 2 Gyr as appears to be required 
to match the PN. Younger stars are brighter and so should
have been well-detected in optical photometry, ruling
out their presence in NGC~5128.

If the PNe studied here belong to this minority younger population of 
age 2-4 Gyr, then the deduced stellar masses, as indicated by the 
HR diagram of the PN central stars in Fig. \ref{HRdiag}, are 
still unexpectedly high. However since these PN belong to the 
bright end of the luminosity function, there is a selection
effect in favour of the more luminous (viz. highest mass) PN
progenitor stars. If however the observed PNe arose from an intermediate 
age population (5-8 Gyr), the progenitor stars would only have a 
mass of $\sim$1.2 M$_\odot$ and the discrepancy between their 
deduced masses and age is more difficult to reconcile with 
evolutionary tracks of single low mass stars (e.g. Vassiliadis 
\& Wood \cite{VasWood}). It would be profitable to study 
spectroscopically a group of lower luminosity PNe in NGC~5128 to 
determine if their properties differ from the high $L$ PNe presented 
in this study.

\section{Conclusions}
Low resolution spectroscopy of 51 PN in NGC~5128 at a range of 
galactocentric distance of 2-20 kpc have been obtained with the 
%ADD
ESO VLT and FORS1 instrument in multi-slit mode (MOS) in three fields. The PN are 
drawn from the upper 4 magnitudes of the PN [O III] luminosity function. The 
emission line spectra have been analysed and lines typical of ionization by 
hot PN central stars have been measured. The weak [O~III]4363\AA\ line 
was just detected in 20\% of the PNe. In order to determine element 
abundances for a larger fraction of the observed PNe, photoionization 
modelling was conducted with Cloudy for 40 PN with the highest quality 
spectra (representing the upper 2 mag. of the PNLF). He, N, O and Ne 
abundances were determined for all the PNe, 
and S and Ar for about half of the sample.
For the most reliably estimated element, oxygen, no
radial gradient is seen and the slope
of O/H v. projected radius is identical inside and outside of 10 kpc.
The range of [O/H] in the sample spanned -0.66 to +0.31 dex 
with the mean 12 + log O/H of 8.52 (median 8.48).
% 8.49 from flux weighted mean.

The PN O abundances were compared to the stellar abundance measured 
from Lick indices from the continuum spectroscopy (i.e. along the 
MOS slitlets not occupied by PN emission) and from resolved star 
photometry from the literature. If the stars in NGC~5128 have an 
average age of 8 Gyr, the stellar and PN metallicities agree in the 
outer parts of NGC~5128 with average values around -0.4 Solar. 
The deduced masses of the PN central stars implies progenitor masses 
above 2 M$_\odot$, favouring their formation from a very young 
component of the intermediate age stellar population with age 
$^{<}_{\sim}$5 Gyr. This discrepancy between the age of the stellar 
population and the mass of the PN progenitor stars is in line with 
other studies of the high mass end of the PN luminosity function.

%Using the original GC calibration of Harris et al. (\cite{Harris}),
%however, the average metallicity from HST photometry and from
%the continuum spectral observations is lower, namely Z=-0.6. The 
%ratio of O from PN and Fe from the stellar indicators is compared with 
%the value found for the possibly comparable environment of the Milky Way
%bulge (from Chiappini et al. \cite{Chiap}).

\begin{acknowledgements}
We would like to thank the staff of Paranal for the very 
efficient conduct of the service observations in programmes 64.N-0219,
66.B-0134, 67.B-0111 and 71.B-0134.

We thank the anonymous referee for many helpful suggestions, including the 
one that led to development of the comparative study of models.
\end{acknowledgements}

\begin{longtable}{l l | r r | r r | r r | r r | r r}
\caption{NGC~5128 PN spectra - Observed line fluxes}
\label{ObsFlux}
\endfirsthead

  &   & \multicolumn{2}{c}{F56 \#1} & \multicolumn{2}{c}{F56 \#2} &  \multicolumn{2}{c}{F56 \#3} & \multicolumn{2}{c}{F56 \#4} & \multicolumn{2}{c}{F56 \#5} \\
  &   & \multicolumn{2}{c}{5615}  & \multicolumn{2}{c}{5601}    &   \multicolumn{2}{c}{5621}   &  \multicolumn{2}{c}{5611}     & \multicolumn{2}{c}{5608} \\
  &   & \multicolumn{2}{c}{($+$297.6,$-$492.5)}  & \multicolumn{2}{c} {($+$284.3,$-$465.9)}  & \multicolumn{2}{c}{($+$287.4,$-$448.7)}  &  \multicolumn{2}{c}{($+$298.7,$-$405.0)}   & \multicolumn{2}{c}{($+$268.9, $-$394.1)} \\
\hline
 Species & $\lambda$ (\AA)  & F$_{Obs}$ & $\pm$ &  F$_{Obs}$ & $\pm$ & F$_{Obs}$ & $\pm$ & F$_{Obs}$ & $\pm$ & F$_{Obs}$ & $\pm$ \\
\hline
~[O~II]                 & 3727 &   38 &  32 &    43 &  10 &   84 &  20 &      &      &      &     \\
~[Ne III]               & 3868 &   38 &  14 &    97 &  10 &   55 &  16 &   72 &  19  &  112 &  60 \\
~[Ne~III] + H$\epsilon$ & 3970 &      &     &    29 &   6 &      &     &      &      &      &     \\
~H$\delta$              & 4101 &      &     &    22 &   8 &      &     &      &      &      &     \\
\hline\hline
\end{longtable}
\newpage

\begin{longtable}{l l | r r | r r | r r | r r | r r}
\caption{NGC~5128 PN spectra - extinction corrected line fluxes}
\label{DeredFlux}
\endfirsthead

  &   & \multicolumn{2}{c}{F56 \#1} & \multicolumn{2}{c}{F56 \#2} &  \multicolumn{2}{c}{F56 \#3} & \multicolumn{2}{c}{F56 \#4} & \multicolumn{2}{c}{F56 \#5} \\
  &   & \multicolumn{2}{c}{5615}  & \multicolumn{2}{c}{5601}    &   \multicolumn{2}{c}{5621}   &  \multicolumn{2}{c}{5611}     & \multicolumn{2}{c}{5608} \\
  &   & \multicolumn{2}{c}{($+$297.6,$-$492.5)}  & \multicolumn{2}{c} {($+$284.3,$-$465.9)}  & \multicolumn{2}{c}{($+$287.4,$-$448.7)}  &  \multicolumn{2}{c}{($+$298.7,$-$405.0)}   & \multicolumn{2}{c}{($+$268.9, $-$394.1)} \\
\hline
 Species & $\lambda$ (\AA)  & F$_{Dered}$ & $\pm$ &  F$_{Dered}$ & $\pm$ & F$_{Dered}$ & $\pm$ & F$_{Dered}$ & $\pm$ & F$_{Dered}$ & $\pm$ \\
\hline
~[O~II]                 & 3727 &   50 &  42 &    56 &  13 &  120 &  29 &      &      &      &     \\
~[Ne III]               & 3868 &   49 &  18 &   124 &  13 &   76 &  22 &   95 &  25  &  150 &  81 \\
~[Ne~III] + H$\epsilon$ & 3970 &      &     &    37 &   8 &      &     &      &      &   72 &  42 \\
~H$\delta$              & 4101 &      &     &    27 &  10 &      &     &      &      &      &     \\
\hline\hline
\end{longtable}

\end{document}